\documentclass[aps, prx, superscriptaddress, twocolumn, nofootinbib]{revtex4-2}

\usepackage[english]{babel}

\usepackage[letterpaper,top=2cm,bottom=2cm,left=2cm,right=2cm,marginparwidth=1.75cm]{geometry}

\usepackage{amsmath}
\usepackage{amssymb }
\usepackage{graphicx}
\usepackage{bm}
\usepackage{braket}
\usepackage{tensor}
\usepackage{mathtools}
\usepackage{dsfont}
\usepackage{subcaption}
\usepackage{mathrsfs}
\usepackage[colorlinks=true, allcolors=blue]{hyperref}

\begin{document}

\title{
% Uncertainty, Correlation, and Entanglement in Quantum Reference Frames: 
Relativity of Quantum Correlations: \\Invariant Quantities and Frame-Dependent Measures
}

\author{Michael Suleymanov}
\email{michael.suleymanov@biu.ac.il}
\affiliation{Faculty of Engineering and the Institute of Nanotechnology and Advanced Materials, Bar-Ilan University, Ramat Gan 5290002, Israel}

\author{Avishy Carmi}
\affiliation{Faculty of Engineering and the Institute of Nanotechnology and Advanced Materials, Bar-Ilan University, Ramat Gan 5290002, Israel}

\author{Eliahu Cohen}
\affiliation{Faculty of Engineering and the Institute of Nanotechnology and Advanced Materials, Bar-Ilan University, Ramat Gan 5290002, Israel}

\begin{abstract}
Viewing frames of reference as physical systems, subject to the same laws as the systems they describe, is central to the relational approach in physics. Under the assumption that quantum mechanics universally governs all physical entities, this perspective naturally leads to the concept of quantum reference frames (QRFs).
We investigate the perspective-dependence of position and momentum uncertainties, correlations, covariance matrices, and entanglement within the QRF formalism. We show that the Robertson-Schr\"odinger uncertainty relations are frame-dependent, and so are correlations and variances, which satisfy various constraints described as inequalities. However, the determinant of the total covariance matrix, linked to the uncertainty volume in phase space, as well as variance-based entanglement criteria, remains invariant under changes of reference frame. Under specific conditions, the purities of subsystems are also invariant for different QRFs, but in general, they are perspective-dependent. These invariants suggest fundamental, robust measures of uncertainty and entanglement that persist despite changes in observational perspective, potentially inspiring dedicated quantum information protocols as well as further foundational studies.

\end{abstract}

\maketitle
% \tableofcontents
\section{Introduction}
Uncertainty relations, introduced by Heisenberg in 1927~\cite{heisenberg1927} and generalized by Robertson~\cite{Robertson1929_Uncertainty_Principle} and Schr\"odinger~\cite{schrodinger1930}, are considered a fundamental feature of quantum mechanics. In nonrelativistic quantum mechanics, position and momentum second moments (variances and correlations) are invariant under Galilean transformations~\cite{zettili_quantum_2022}. 
Another cornerstone of quantum mechanics is the notion of entanglement, whose conceptual intricacy was introduced by Einstein, Podolsky and Rosen in~\cite{EPR1935} and Schr\"odinger in~\cite{schrodinger1935discussion,schrodinger1936Probability}.
The main claim of the relational approach is that the description of physical objects has a meaning only from the perspective of another physical object, serving in that case as a reference frame. In such a framework, frames of reference are no longer abstract, they have an internal structure and might have correlations and interactions with the described objects. Assuming that nature is quantum on the fundamental level, physical systems that serve as frames of reference, as well as the described ones, should be treated as quantum. The framework that follows these concepts, called quantum reference frames,  
was first introduced in~\cite{aharonov1967observability,Aharonov_Susskind_ssr,Aharonov_Kaufherr_qrf} and further developed in
\cite{
% General
Rovelli_1991,
Kitaev_2004,
Angelo_2011,
Yang2020switchingquantum,
pereira2015galilei,
PhysRevA.103.022220,
% Informational
bartlett2006degradation,
Bartlett2007,
bartlett2009quantum,
Gour_Spekkens2008resource,
girelli2008quantum,
smith2016quantum,
castro2021relative,
martinelli2019quantifying,
% Operational
Miyadera_2016,
Loveridge_2017,
Loveridge_2018,
riera2024uncertaintyrelationsrelativephase,
% Perspectival
Giacomini_2019,
giacomini2019relativistic,
de_la_Hamette_2020,
cepollaro2024sumentanglementsubsystemcoherence,
% Perspective-Neutral
vanrietvelde2018switching,
2020_change_of_perspective,
hohn2020switch,
de2021perspective,
ahmad2022quantum,
giacomini2021spacetime,
suleymanov2023nonrelativistic_PRA,
hoehn2023quantum,
% Subsystems Relativity
ahmad2022quantum,
de2021perspective,
castro2021relative,
hoehn2023quantum,
% Correlations perspective dependence
% Gravitational entropy
devuyst2024gravitationalentropyobserverdependent,
devuyst2024crossedproductsquantumreference,
ahmad2024relationalquantumgeometry,
% Moments and UR
Bojowald_2011_1,
Bojowald_2011_2,Hohn_2012}.
Several approaches to quantum reference frames have been developed in recent years.
The 
\textit{informational} approach~\cite{Bartlett2007,bartlett2009quantum,Gour_Spekkens2008resource,bartlett2006degradation,girelli2008quantum,smith2016quantum,castro2021relative,martinelli2019quantifying}  views quantum reference frames as information carriers and emphasizes information-theoretic principles as fundamental.
The \textit{operational} approach~\cite{Miyadera_2016,Loveridge_2017,Loveridge_2018,riera2024uncertaintyrelationsrelativephase}, 
defined through explicit measurement procedures or laboratory protocols, focuses on observable consequences and explicit experimental methods for establishing and comparing frames.
In the \textit{perspectival} approach~\cite{Giacomini_2019,giacomini2019relativistic,de_la_Hamette_2020,cepollaro2024sumentanglementsubsystemcoherence}, quantum states and dynamics explicitly depend on the chosen perspective while in \textit{perspective-neutral} formalism~\cite{vanrietvelde2018switching,2020_change_of_perspective,hohn2020switch,de2021perspective,ahmad2022quantum,giacomini2021spacetime,suleymanov2023nonrelativistic_PRA,hoehn2023quantum}, the existence of a kinematical state is assumed before the choice of QRF.
% {\color{blue} The approach that we call here as \textit{perspectival} is sometimes called \textit{operational} in the literature. We distinguish } 
The relativity of subsystems was first described in~\cite{ahmad2022quantum} and further studied in~\cite{de2021perspective,castro2021relative,hoehn2023quantum}. 
The perspective dependence of quantum correlations and entanglement has been studied from various viewpoints: 
subsystem relativity of gravitational entropy in~\cite{devuyst2024gravitationalentropyobserverdependent,devuyst2024crossedproductsquantumreference,ahmad2024relationalquantumgeometry}, 
second moments and uncertainties in~\cite{Bojowald_2011_1,Bojowald_2011_2,Hohn_2012}, subsystem relativity of entropy and thermodynamics in~\cite{hoehn2023quantum}, 
the sum of entanglement and coherence in~\cite{cepollaro2024sumentanglementsubsystemcoherence}.

\vspace{1mm}

In this work, we concentrate on uncertainties, correlations and entanglement in the \textit{perspective-neutral} approach to QRFs~\cite{2020_change_of_perspective}, and explore how the fact that frames of reference are considered to be physical (and quantum) objects changes our understanding of these concepts. We show that covariances associated with individual particles are perspective-dependent, but the ones corresponding to the entire system, and specifically, the determinants of the total covariance matrices, are invariant under QRF transformations. Throughout the work, we intensively use the unique structure of the covariance matrix, whose significance within and beyond quantum theory was studied, for instance, in~\cite{Carmi_2018,Carmi_2019, Rel_indep_bounds}. Additional invariant quantities are the variance-based continuous-variable entanglement criteria, introduced in \cite{Reid1989,Duan2000,Simon2000}.

In relation to the study of second moments and correlations within QRFs, we would like to briefly address a few recent works. Riera and Loveridge rigorously study the position-momentum uncertainty relations relative to a phase space QRF in the formalism introduced in~\cite{riera2024uncertaintyrelationsrelativephase}, and investigate the classical limit in a Galilei-invariant setting.
Here we explore all the second moments individually, and the covariance matrix as a whole, concentrating on the QRF transformations. 
In a comprehensive work by H\"ohn, Kotecha and Mele~\cite{hoehn2023quantum}, the authors expand the QRF relativity of subsystems investigating how correlations, entropies, and thermodynamical processes change under switching perspectives. Among other results, it was shown that entanglement entropies of subsystems are invariant under QRF transformations if the global state is in a gauge-invariant subalgebra that is such that QRF transformations act on it as if they were multilocal unitary states.
% {\color{red} In the current paper, the reciprocal entanglement entropy is shown to be invariant under switching perspectives in the perspective-neutral QRF formalism~\cite{2020_change_of_perspective}, in the case of Gaussian states. } 
The invariance of an entanglement-related quantity under switching QRFs was discussed in another alluring paper~\cite{cepollaro2024sumentanglementsubsystemcoherence} by Cepollaro, Akil, Cie\'sli\'nski, de la Hamette and Brukner. 
% In this paper, the authors consider three two-level systems and show that the sum of entanglement entropy and subsystem coherence is conserved under QRF transformation. 
In this paper, the authors investigate a bipartite system from the perspective of another system, serving as a reference frame, within the perspectival QRF formalism~\cite{Giacomini_2019}. They demonstrate that the sum of entanglement entropy and subsystem coherence is conserved under a general QRF transformation for any locally compact group and associated Hilbert space.
In the current study, among other results, we take a different route exploring entanglement in QRF formalism by examining the variance-based entanglement criteria~\cite{Reid1989,Duan2000,Simon2000}, which is shown to be perspective-independent. 
% entropy, considering a continuous-variable $\mathscr{N}$-particle system, prepared in a Gaussian state. Such states are completely characterized by the corresponding covariance matrices, whose QRF transformations are at the core of the present work. Expressing the reduced density matrices in terms of the covariance matrices, we show 
% that the entanglement entropy and the entropy of coherence are perspective-invariant independently in our case. 
% While our discussion of entanglement is restricted to a particular class of states, it extends the analysis to a system with an arbitrary number of particles.
Consequently, the scenario explored here, and the resulting conclusions, differ from those presented in~\cite{hoehn2023quantum} and~\cite{cepollaro2024sumentanglementsubsystemcoherence}.\\
% \vspace{1mm}

The structure of this work is as follows.
In Sec.~\ref{Sec_Methods} we provide a brief introduction to the perspective-neutral approach to quantum reference frames and present the main notations used in the current manuscript. We also formulate the perspective-dependent second moments, which are the core ingredients in the current work. 
The main results are presented in Sec.~\ref{Sec_Results}. 
We start with showing the perspective-dependence of position and momentum second moments and then develop relations and bounds between them in different QRFs in Sec.~\ref{sec:QRF_sec_mom}. 
The divergence of the purities of subsystems, and their cohesion under specific conditions, in different QRFs, in the case of Gaussian states, is discussed in Sec.~\ref{sec:QRF_sec_purity}.
Next, in Sec.~\ref{Sec_invariants}, we show that the determinant of the total position-momentum covariance matrix is invariant under changing perspectives, as well as the variance-based entanglement criteria. In the following Sec.~\ref{Sec_Time_evol} the time evolution is considered. First, we show the deviation of the position-momentum uncertainty relation expression in the non-interacting case, compared to the non-relational framework even without changing perspectives. Second, exploring systems governed by quadratic Hamiltonian, we find that the determinant of the total position-momentum covariance matrix in this case is conserved concerning time evolution. We conclude our work in Sec.~\ref{Discussion_Conclussions}.\\ 
% \vspace{1mm}

The main novelty of the current paper is the rigorous study of perspective-dependence of uncertainties, correlations, and entanglement, as well as their interrelations, while introducing and analyzing new invariant quantities in the QRF formalism. 
% which are perspective-neutral.

\section{The formalism}\label{Sec_Methods}
% \subsection{Spatial quantum reference frames}
% \label{sec:QRF_intro}
% \vspace{1cm}
% The concept of the relational approach in quantum mechanics was first introduced by Aharonov and Susskind in 1967 \cite{Aharonov_Susskind_ssr}, and gained a wide interest since then {\red [refs]}...
In the current manuscript we use the spatial quantum reference frames, based on and motivated by~\cite{2020_change_of_perspective}. Here we will present the key ideas.

\begin{figure}[h]
\centering
\includegraphics[width=0.45\textwidth]{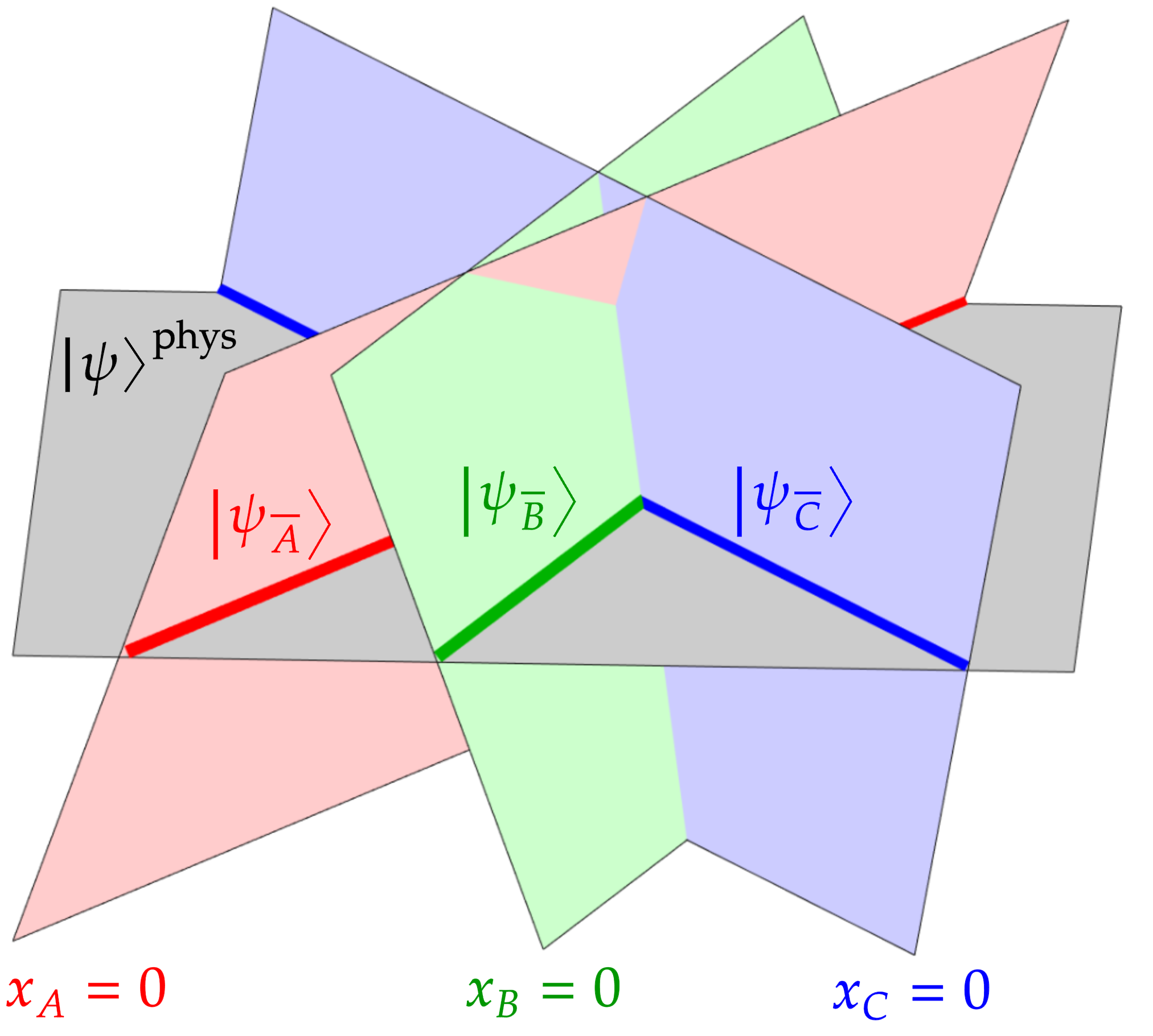}
\caption{The intersections between the physical state and the choice of QRF constraints in Eq.~\eqref{eq_QRF_choice}, reflect the perspective-dependent descriptions.
}
\label{FIG:QRF_choice}
\end{figure}

The total, kinematical, Hilbert space is constructed as a tensor product of Hilbert spaces associated with $\mathscr{N}$ individual objects (particles) $\mathcal{H}^\text{kin}=\bigotimes_I\mathcal{H}_I$ with states $\ket{\Psi}^\text{kin}\in \mathcal{H}^\text{kin}$ 
where $I\in\mathfrak{N}$ and $\mathfrak{N}\coloneqq\{A,...,\mathscr{N}\}$. 
An arbitrary state in $\mathcal{H}^{\text{kin}}$ can be written as
    \begin{equation}
    \begin{aligned}
\ket{\Psi}^{\text{kin}} =
&\int \prod_{I\in\mathfrak{N}}  dp_I \ket{p_I}
\Psi^\text{kin}(p_A,...,p_\mathscr{N}) =\\
&\int  dP \ket{P}  \Psi^\text{kin}(P), 
    \end{aligned}
    \end{equation}
where we have introduced $P \coloneqq \{p_I; \; I \in \mathfrak{N}\}$ and $dP=\prod_{I\in\mathfrak{N}}dp_I$.
The total Hamiltonian, for the non-relativistic case, is of the form, 
\begin{equation}
\hat{H}_\text{tot}=\sum_{I\in\mathfrak{N}}\frac{\hat{p}_I^2}{2m_I}+V(\{\hat{x}\}),
\end{equation}
where $V(\{\hat{x}\})=V(\hat{x}_A,...,\hat{x}_\mathscr{N})$. 
Physical states, $\ket{\Psi}^\text{phys}$, satisfying the translation invariance constraint,
\begin{equation}
    \hat{P}_\text{tot}\ket{\Psi}^\text{phys}=0,
    \label{eq:momentum_constraint}
\end{equation}
where,
$\hat{P}_\text{tot}=\sum_{I\in\mathfrak{N}}\hat{p}_I$, 
are obtained using what is known as group-averaging $\ket{\psi}^\text{phys}=\delta(\hat{P}_\text{tot})\ket{\psi}^\text{kin} =\int dP  \delta(P_\text{tot})
\ket{P}\Psi^\text{kin}(P)\in\mathcal{H}^\text{phys}$. 
In the perspective-dependent view, say $A$'s, it takes the form,
% In the momentum representation it may be written as, $\ket{\Psi}^{\text{phys}}=
% \int dP  \delta(P_\text{tot})
% \ket{P}\Psi^\text{kin}(P)$ or, in perspective-dependent view, say $A$'s, 
\begin{equation}
\begin{aligned}
&\ket{\Psi}^{\text{phys}} =\\
&\int \ket{p_A=-p_{\bar A}}\prod_{I\in\mathfrak{A}} dp_I \ket{p_I}
\Psi^\text{kin}(p_A=-p_{\bar A},...,p_\mathscr{N}),  
\end{aligned}
\end{equation}
where $\mathfrak{A}\equiv\mathfrak{N}\setminus \{A\}$, and,
\begin{equation}\begin{aligned}p_{\bar A}=\sum_{I\in\mathfrak{A}}p_I.
\label{eq:p_bar_J}
\end{aligned}\end{equation}
Denoting 
$\psi_{\bar A}(P_{\bar A})=
\Psi^\text{kin}(p_A=-p_{\bar A},...,p_\mathscr{N})$, 
$dP_{\bar A}=\prod_{I\in\mathfrak{A}}  dp_I$, and  $\ket{P_{\bar A}}=\prod_{I\in\mathfrak{A}}\ket{p_I}$, we may write,
\begin{equation}
    \begin{aligned}
\ket{\Psi}^{\text{phys}} &=
\int dP_{\bar A}\ket{p_A=-p_{\bar A}}\ket{P_{\bar A}} \psi_{\bar A}.
    \end{aligned}
    \end{equation}
For any integrable function $f(P)$ and for every $I,J\in\mathfrak{N}$, it holds that
\begin{equation}
    \label{eq:identity-any-frame}
    \begin{aligned}
    \int  dP& \delta(P_T) f(P) 
    =\int dP_{\bar I} f(p_I=-p_{\bar I},P_{\bar I}) =\\
    &\int dP_{\bar I} f_{\bar I}(P_{\bar I})=\int dP_{\bar J} f_{\bar J}(P_{\bar J}).
    \end{aligned}
\end{equation}
Following the procedure discussed in \cite{2020_change_of_perspective} the state from a perspective of particle $I$, as illustrated in Fig.~\ref{FIG:QRF_choice}, is of the form,
\begin{equation}
\ket{\psi_{\bar I}} \coloneqq \sqrt{2\pi} \Braket{x_I=0|\Psi}^{\text{phys}}=
\int dP_{\bar I }\ket{P_{\bar I}} \psi_{\bar I}.
\label{eq_QRF_choice}
\end{equation}
It can be shown~\cite{2020_change_of_perspective} that the perspective-dependent Hamiltonian of the form
\begin{equation}
\hat{H}_{\bar A}=\frac{\hat{p}_{\bar A}^2}{2m_A} +
\sum_{I\in\mathfrak{A}}\frac{\hat{p}_I^2}{2m_I}+V_{\bar A},
\label{eq:Hamiltonian_persp_dep}
\end{equation}
where $V_{\bar A}=\Big[V(\{x\})\Big]_{x_A=0}$, governs the dynamics of $\ket{\psi_{\bar A}}$, i.e.,
\begin{equation}
    i\frac{d}{dt}\ket{\psi_{\bar A}(t)} = \hat{H}_{\bar A} \ket{\psi_{\bar A}(t)},
\end{equation}
where $\ket{\psi_{\bar A}(0)} = \ket{\psi_{\bar A}}$, and, $\ket{\psi_{\bar A}(t)}=
\int dP_{\bar I} e^{-itH_{\bar A}} \ket{P_{\bar A}}\psi_{\bar A}$.
As discussed in~\cite{2020_change_of_perspective}, in the limit when $m_A\gg m_{I\in \mathfrak{N}\setminus\{A\}}$, the non-relational Hamiltonian is recovered with respect to an abstract reference frame. In the following, we will show that it is not true for the QRF transformations of the second moments.\\

The expression for the expectation value of an operator $\hat O$ (that does not depend explicitly on time) from $I$'s perspective is of the form,
\begin{equation}
\begin{aligned}
\langle\hat{O}\rangle_I(t)=
&\langle\psi_{\bar I}(t)|\hat{O}|\psi_{\bar I}(t)\rangle=\\
&\langle\psi_{\bar I}(t=0) e^{it\hat{H}_{\bar I}}
|\hat{O}|
e^{-it\hat{H}_{\bar I}} \psi_{\bar I}(t=0)\rangle.
\end{aligned}
\end{equation}
Similarly, the expressions for variances and covariances are respectively of the form,
\begin{equation}
    \sigma^2_{I}(\hat{O})=
    \braket{\hat{O}^2}_I -
    \braket{\hat{O}}_I^2,
\end{equation}
and 
\begin{equation}
\begin{aligned}
&\text{cov}_{I}(\hat{O}_1,\hat{O}_2)=
\text{cov}_{I}(\hat{O}_2,\hat{O}_1)=\\
&\frac{1}{2}\left(\braket{\hat{O}_1\hat{O}_2}_I+\braket{\hat{O}_2\hat{O}_1}_I\right) -
\braket{\hat{O}_1}_I\braket{\hat{O}_2}_I.
\end{aligned}
\end{equation}
When it is sufficiently clear, we omit the dependency on time.\\ 

The covariance matrix associated with $\hat{O}_1$ and $\hat{O}_2$, from $I$'s perspective,
is defined as follows
\begin{equation}
    \begin{aligned}
\bm{\Sigma}_{(I)}^{(\hat{O}_1,\hat{O}_2)}=
\left[\begin{matrix}
\sigma_I^2(\hat{O}_1)&\text{cov}_I(\hat{O}_1,\hat{O}_2)\\
\text{cov}_I(\hat{O}_2,\hat{O}_1)&\sigma_I^2(\hat{O}_2)
\end{matrix}\right].
    \end{aligned}
\end{equation}

\section{Invariant Quantities and Frame-Dependent Measures}\label{Sec_Results}
% \section{Uncertainties and correlations in QRFs at the initial instant}\label{sec:uncertainty relations in QRFs}
% \subsection{Spatiotemporal quantum reference frames}

\subsection{Relations between position and momentum variances and correlations in different QRFs at the initial instant}
\label{sec:QRF_sec_mom}
Since we are focusing on the initial instant in this section, when the Hamiltonian doesn't play a role, the results are quite general and relevant for all systems obeying the translation invariance constraint, including relativistic scenarios~\cite{giacomini2021spacetime}.
The transformations between different QRFs of the second position and momentum moments are presented in Appendix~\ref{app_sec_EV_UR}. 
The relations between the second moments for the initial instance hold for any $t$ in the non-interacting case.
The position second moments transformations between different QRFs may be obtained using Eq.~\eqref{app:eq_trans_f(x)}.
The reciprocal position variances of any two QRFs coincide,
\begin{equation}\begin{aligned}
\sigma^2_I\left(\hat{x}_J\right)=
\sigma^2_J\left(\hat{x}_I\right),
\label{eq:var_x_reciprocal}
\end{aligned}\end{equation} 
when their description of other particles does not. To see this, we need the QRF transformations of the position moments derived in Eqs.~\eqref{app:eq_var_cov_x_1},~\eqref{app:eq_cov_x_1} and~\eqref{app:eq_cov_x_2},
% and~\eqref{app:eq_cov_x_3}, 
% as opposed to Eqs.~\eqref{eq_UR_non-rel} and~\eqref{eq_cov_non-rel},
\begin{equation}
\begin{aligned}
\sigma^2_I(\hat{x}_K)=
\sigma^2_J(\hat{x}_I)+\sigma^2_J(\hat{x}_K)-
2\text{cov}_J(\hat{x}_I,\hat{x}_K)
\label{eq_var_cov_x},
\end{aligned}
\end{equation} 
\begin{equation}
\begin{aligned}
\text{cov}_I(\hat{x}_J,\hat{x}_K)=
-\text{cov}_J(\hat{x}_I,\hat{x}_K)+
\sigma^2_J(\hat{x}_I),
\end{aligned}
\label{eq_cov_cov_var_x}
\end{equation}
\begin{equation}
\begin{aligned}
\text{cov}_I(\hat{x}_J,\hat{x}_K) = 
&\text{cov}_L(\hat{x}_J,\hat{x}_K)-
\text{cov}_L(\hat{x}_J,\hat{x}_I)-\\
&\text{cov}_L(\hat{x}_K,\hat{x}_I)+
\sigma^2_L(\hat{x}_I),
\end{aligned}
\label{eq_cov_x_2}
\end{equation}
where $I,J,K,L\in\mathfrak{N}$. 
According to Eq.~\eqref{eq:var_x_reciprocal} we can write the difference between the position variances of a certain particle, say $K$, from different perspectives, say $I$'s and $J$'s,
\begin{equation}\begin{aligned}
\sigma_I^2(\hat{x}_K)-\sigma_J^2(\hat{x}_K)=
\sigma_K^2(\hat{x}_I)-\sigma_K^2(\hat{x}_J).
\end{aligned}\end{equation} 
This means that if the particles $I$ and $J$, have the same variance from $K$'s perspective, then the variances of particle $K$ in their QRFs coincide. \\

Using Eq.~\eqref{eq_var_cov_x} we obtain the relation between the covariance of two particles from different perspectives,
\begin{equation}
    \begin{aligned}
&\text{cov}_I(\hat{x}_K,\hat{x}_L)-
\text{cov}_J(\hat{x}_K,\hat{x}_L)=\\
&\frac{1}{2}\left[
\left(\sigma_I^2(\hat{x}_K)-\sigma_J^2(\hat{x}_K)\right)+
\left(\sigma_I^2(\hat{x}_L)-\sigma_J^2(\hat{x}_L)\right)
\right].
\label{eq_var_cov_x_2}
    \end{aligned}
\end{equation}
The relation between the position variances of a certain particle from different perspectives is obtained from Eq.~\eqref{eq_cov_cov_var_x},
\begin{equation}
    \begin{aligned}
\sigma_I^2(\hat{x}_K)-\sigma_J^2(\hat{x}_K)=
&\text{cov}_I(\hat{x}_J,\hat{x}_K)-\text{cov}_J(\hat{x}_I,\hat{x}_K).
\label{eq_var_x_QRF_dep}
    \end{aligned}
\end{equation}
We can see that the position second moments in Eqs.~\eqref{eq_var_cov_x_2} and~\eqref{eq_var_x_QRF_dep} are perspective-dependent in the QRF formalism, as opposed to the non-relational case in Eqs.~\eqref{eq_UR_non-rel} and~\eqref{eq_cov_non-rel}.

By substituting Eq.~\eqref{eq_cov_cov_var_x} into Eq.~\eqref{eq_cov_x_2}, as we show in Eq.~\eqref{app:eq_cov_x_3}, one gets,
\begin{equation}
\begin{aligned}
&\text{cov}_I(\hat{x}_K,\hat{x}_L)-
\text{cov}_J(\hat{x}_K,\hat{x}_L)=\\
&\text{cov}_I(\hat{x}_J,\hat{x}_K)-
\text{cov}_J(\hat{x}_I,\hat{x}_L)=\\
&\text{cov}_I(\hat{x}_J,\hat{x}_L)-
\text{cov}_J(\hat{x}_I,\hat{x}_K),
\end{aligned}
\label{eq_cov_x_3}
\end{equation}
which relates the covariances from the perspectives of any two QRFs, say $I$'s and $J$'s, considering any two particles, say $K$ and $L$.

It is worthwhile to mention here that in the limit, discussed in~\cite{2020_change_of_perspective}, when the masses associated with QRFs, $I$ and $J$, are much bigger than the ones of the described particles, $m_{I,J}\gg m_{a\in \mathfrak{N}\setminus\{I,J\}}$, the position second moments from different perspectives in Eqs.~\eqref{eq_var_cov_x_2} and~\eqref{eq_var_x_QRF_dep} still do not coincide with the non-relational description.\\

For any three particles $I,J,K\in\mathfrak{N}$, illustrated in Fig.~\ref{FIG2a_x_moments}, the position second moments, from each other's perspectives, as we show in Appendix~\ref{app_subsec_position}, are equivalent to a Euclidean triangle, illustrated in Fig.~\ref{FIG2b_x_triangle}. 
In this analogy, the variances, $\sigma_a(\hat{x}_b)$, represent the side lengths, and the opposite angles, $\alpha_c(\hat{x}_a,\hat{x}_b)$, are related to the correlations, $\cos(\alpha_a(\hat{x}_b,\hat{x}_c))=\text{corr}_a(\hat{x}_b,\hat{x}_c)$, where $a\ne b\ne c\in\left\{I,J,K\right\}$. 

\begin{figure}[h]
\centering
\begin{subfigure}{5cm}
\centering
\includegraphics[width=1\textwidth]{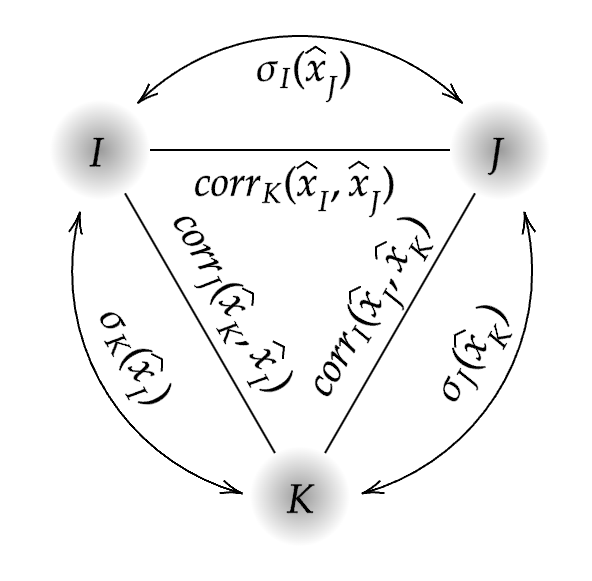}
\caption{A schematic representation of the position variances and correlations of any three particles, $I,J,K\in\mathfrak{N}$, in each other's QRFs.}
\label{FIG2a_x_moments}
\end{subfigure}
\begin{subfigure}{5cm}
\centering
\includegraphics [width=1\linewidth]
{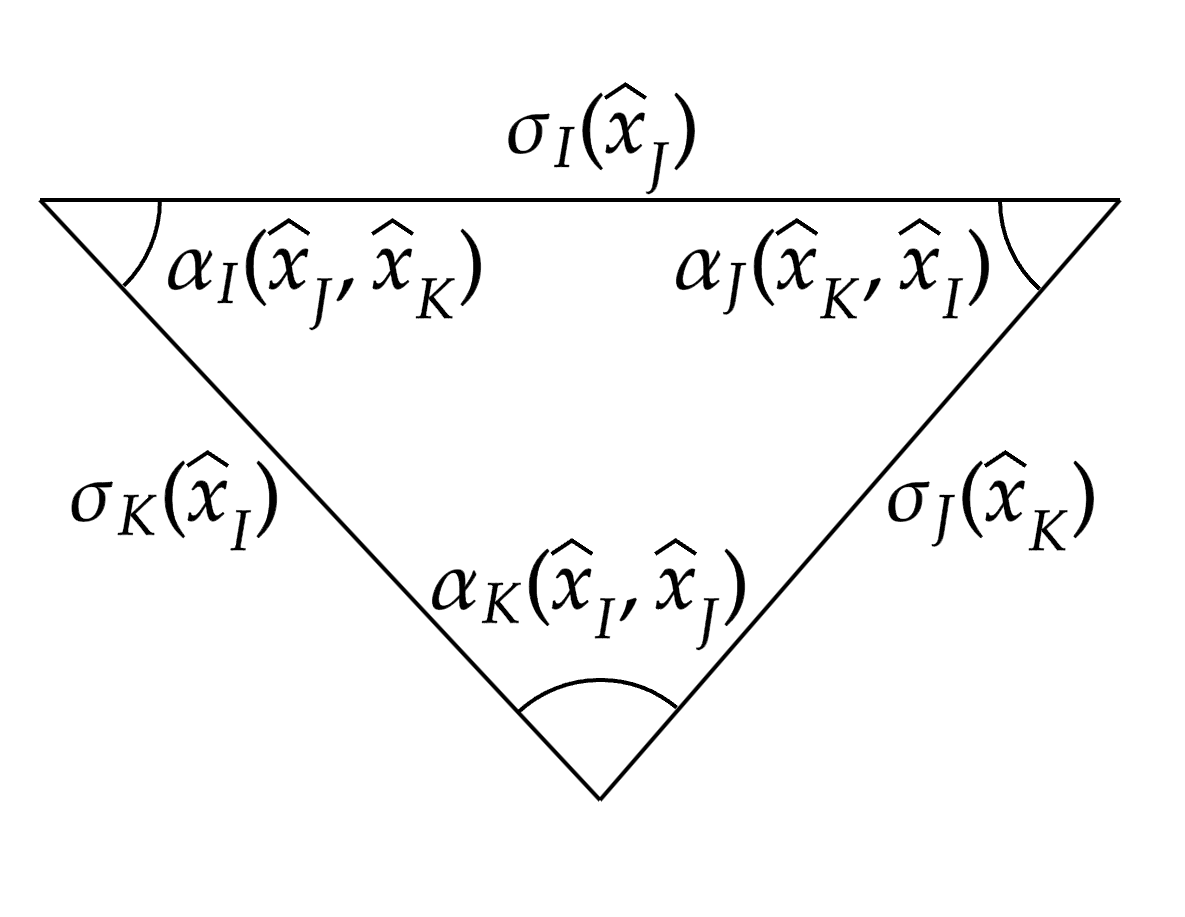}
\caption{The triangle equivalence in which the variances $\sigma_a(\hat{x}_b)$ correspond to the side lengths, and the correlations represent the cosines: $\cos(\alpha_a(\hat{x}_b,\hat{x}_c))=\text{corr}_a(\hat{x}_b,\hat{x}_c)$.}
\label{FIG2b_x_triangle}
\end{subfigure}
% \begin{subfigure}{6cm}
% \centering
% \includegraphics[width=1\linewidth]{FIG2c_x_corr_ineq_3D.png}
% \caption{The forbidden region (in gray) for the position correlations.}
% \label{FIG2c_x_corr_ineq_3D}
% \end{subfigure}
\caption{Relations between the position second moments of any three particles, $I,J,K\in\mathfrak{N}$, describing each other.}
\label{FIG2_position_moments}
\end{figure}

From Eq.~\eqref{eq_var_cov_x}, which may be identified as the cosine law in this equivalence,
\begin{equation}
\begin{aligned}
&\sigma^2_a(\hat{x}_b)=\\
&\sigma^2_c(\hat{x}_a)+\sigma^2_c(\hat{x}_b)-
2\sigma_c(\hat{x}_a)\sigma_c(\hat{x}_b)
\cos(\alpha_a(\hat{x}_b,\hat{x}_c)),
\end{aligned}
\end{equation}
we obtain that the position variances obey the triangle inequalities,
\begin{equation}\begin{aligned}
|\sigma_c(\hat{x}_a)-\sigma_c(\hat{x}_b)|
\le
\sigma_a(\hat{x}_b)
\le 
\sigma_c(\hat{x}_a)+\sigma_c(\hat{x}_b).
\label{eq_var_x_ineq}
\end{aligned}\end{equation} 
As we show in Appendix~\ref{app_subsec_position}, the correlations satisfy the inequalities, (Eq.~\eqref{app:eq_corr_ineq_2})
\begin{equation}
\text{corr}_a(\hat{x}_b,\hat{x}_c)
\text{corr}_b(\hat{x}_c,\hat{x}_a)+
\text{corr}_c(\hat{x}_a,\hat{x}_b)\ge0,
\end{equation}
and (Eq.~\eqref{app:eq_corr_ineq_x_1})
\begin{equation}
\text{corr}_a(\hat{x}_b,\hat{x}_c)+\text{corr}_d(\hat{x}_e,\hat{x}_f)\ge0,
\end{equation}
where $d\ne e\ne f\in\left\{I,J,K\right\}$. In addition, the constraint in Eq.\eqref{app_eq_x_corr_constraint},
\begin{equation}
    \begin{aligned}
    &\text{corr}_I^2(\hat{x}_J,\hat{x}_K)+
    \text{corr}_J^2(\hat{x}_K,\hat{x}_I)+
    \text{corr}_K^2(\hat{x}_I,\hat{x}_J)
    +\\
    &2\text{corr}_I(\hat{x}_J,\hat{x}_K)
    \text{corr}_J(\hat{x}_K,\hat{x}_I)
    \text{corr}_K(\hat{x}_I,\hat{x}_J)
    =1,
    \end{aligned}
\end{equation}
imposes bounds on the inequalities governing the correlations' sum 
(Eq.\eqref{app_eq_x_corr_sum}),
\begin{equation}
    1\le\text{corr}_I(\hat{x}_J,\hat{x}_K)+
    \text{corr}_J(\hat{x}_K,\hat{x}_I)+
    \text{corr}_K(\hat{x}_I,\hat{x}_J)\le 3/2,
\end{equation}
and product (Eq~\eqref{app_eq_x_corr_prod}),
\begin{equation}
    -1\le \text{corr}_I(\hat{x}_J,\hat{x}_K)
    \text{corr}_J(\hat{x}_K,\hat{x}_I)
    \text{corr}_K(\hat{x}_I,\hat{x}_J)\le 1/8.
\end{equation}

\begin{figure}[h]
\centering
\begin{subfigure}{7cm}
\centering
\includegraphics[width=1\textwidth]{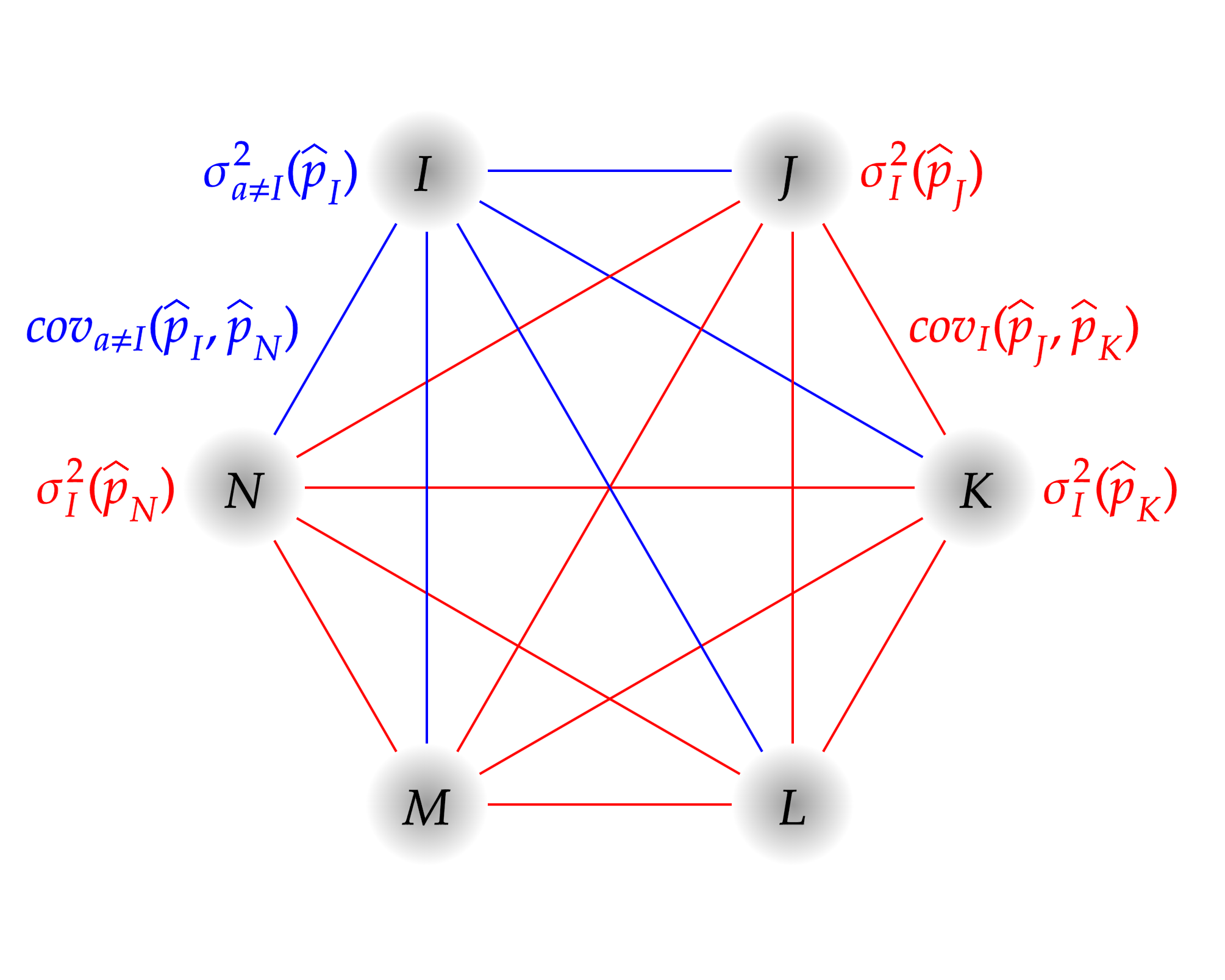}
\caption{Momentum second moments from the perspective of system $I$ (in red), as well as those associated with $I$ from other perspectives (in blue), as appears in Eq.~\eqref{eq:sigma(p)_transf}.
}
\label{FIG3a_p_moments_1}
\end{subfigure}
\begin{subfigure}{7cm}
\centering
\includegraphics[width=1\linewidth]{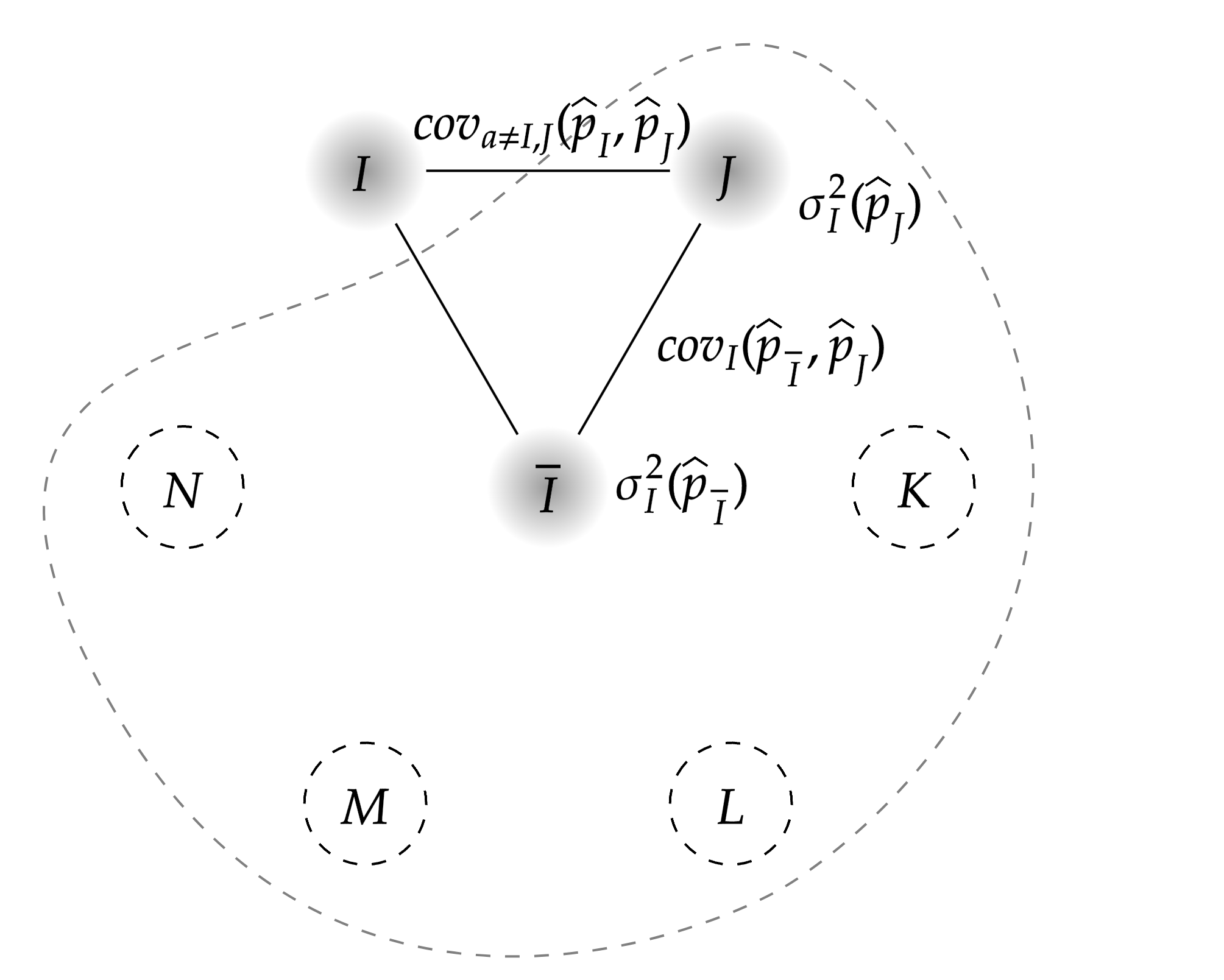}
\caption{Momentum variance of particle $J$ and its covariance with the rest of the system, $\bar I$ (including $J$), from $I$'s perspective, as appears in Eq.~\eqref{eq:cov_p_trnas}.}
\label{FIG3b_p_moments_2}
\end{subfigure}
\caption{Relations between the momentum second moments in different QRFs.}
\label{FIG3_momentum_moments}
\end{figure}

The expectation values related to the momentum degrees of freedom, according to Eq.~\eqref{eq:<p_I>_any_short}, coincide for all QRFs,
\begin{equation}
    \begin{aligned}
\sigma^2_I(\hat{p}_K)=\sigma^2_J(\hat{p}_K),
    \end{aligned}
\label{eq:var_p_inv}
\end{equation}
and,
\begin{equation}
\text{cov}_{I}(\hat{p}_K,\hat{p}_L)=
\text{cov}_{J}(\hat{p}_K,\hat{p}_L).
\label{eq:cov_p_inv}
\end{equation}
But, in the reciprocal transformations, where the new QRF is the one that is described in the previous one, the situation is different, as we show in Eq.~\eqref{app:eq_trans_f(p)}.
The transformation of the momentum variance of particle $I$, illustrated in Fig.~\ref{FIG3a_p_moments_1}, according to Eq.~\eqref{app_eq_sigma(p)_transf}, is of the form,
\begin{equation}
    \begin{aligned}
\sigma^2_{a\ne I}(\hat{p}_I)=
&\sigma^2_{I}(\hat{p}_{\bar I})=
\sum_{b,c \ne I}\text{cov}_I(\hat{p}_b,\hat{p}_c)=\\
&\sum_{b\ne I}\sigma^2_I(\hat{p}_b)+
\sum_{b\ne c \ne I}\text{cov}_I(\hat{p}_b,\hat{p}_c),
    \end{aligned}
    \label{eq:sigma(p)_transf}
\end{equation}
where $-\hat{p}_{\bar I}$ is the total momentum from $I$' perspective, as defined in 
Eq.~\eqref{eq:p_bar_J}. 

The momentum-covariance between particles $I$ and $J$,  illustrated in Fig.~\ref{FIG3b_p_moments_2}, according to Eq.~\eqref{app:eq_cov_p_trnas}, transforms as follows,
\begin{equation}
\begin{aligned}
&\text{cov}_{a\ne I,J}(\hat{p}_I,\hat{p}_J)=
-\text{cov}_{b}(\hat{p}_{\bar b},\hat{p}_c)=\\
&-\sigma_b^2(\hat{p}_c)
-\sum_{d\ne I,J}\text{cov}_b(\hat{p}_d,\hat{p}_c)=\\
&-\sigma_b^2(\hat{p}_c)-
\text{cov}_b(\hat{p}_{\overline{IJ}},\hat{p}_c),
\end{aligned}
\label{eq:cov_p_trnas}
\end{equation}
where $b,c=I,J$ and $\hat{p}_{\overline{IJ}}
=\sum_{K\ne I,J}\hat{p}_K$, 
is related to the variance of $J$ and the sum of its covariances with the rest of the system from the perspective of $I$, and vice versa, illustrated in Fig.~\ref{FIG4a_p_moments}. 
As we show in Appendix~\ref{app_subsec_momentum}, the inequalities regarding the momentum variances and correlations, are of the form,
\begin{equation}\begin{aligned}
|\sigma_I(\hat{p}_J)-\sigma_J(\hat{p}_I)|\le\sigma_{I,J}(\hat{p}_{\overline{IJ}}),
\label{eq:var_p_ineq}
\end{aligned}\end{equation} 
and,
\begin{equation}\begin{aligned}
&
\text{corr}_a(\hat{p}_{ \overline{IJ}},\hat{p}_b)
\le 
\text{corr}_b(\hat{p}_{\overline{IJ}},\hat{p}_a)
\text{corr}_{c\ne I,J}(\hat{p}_a,\hat{p}_b),
\label{eq_p_corr_ineq}
\end{aligned}
\end{equation} 
illustrated in Fig.~\ref{FIG4b_p_corr_ineq_3D}, where $a,b=I,J$.
\begin{figure}[h]
\centering
\begin{subfigure}{7cm}
\centering
\includegraphics[width=1\textwidth]{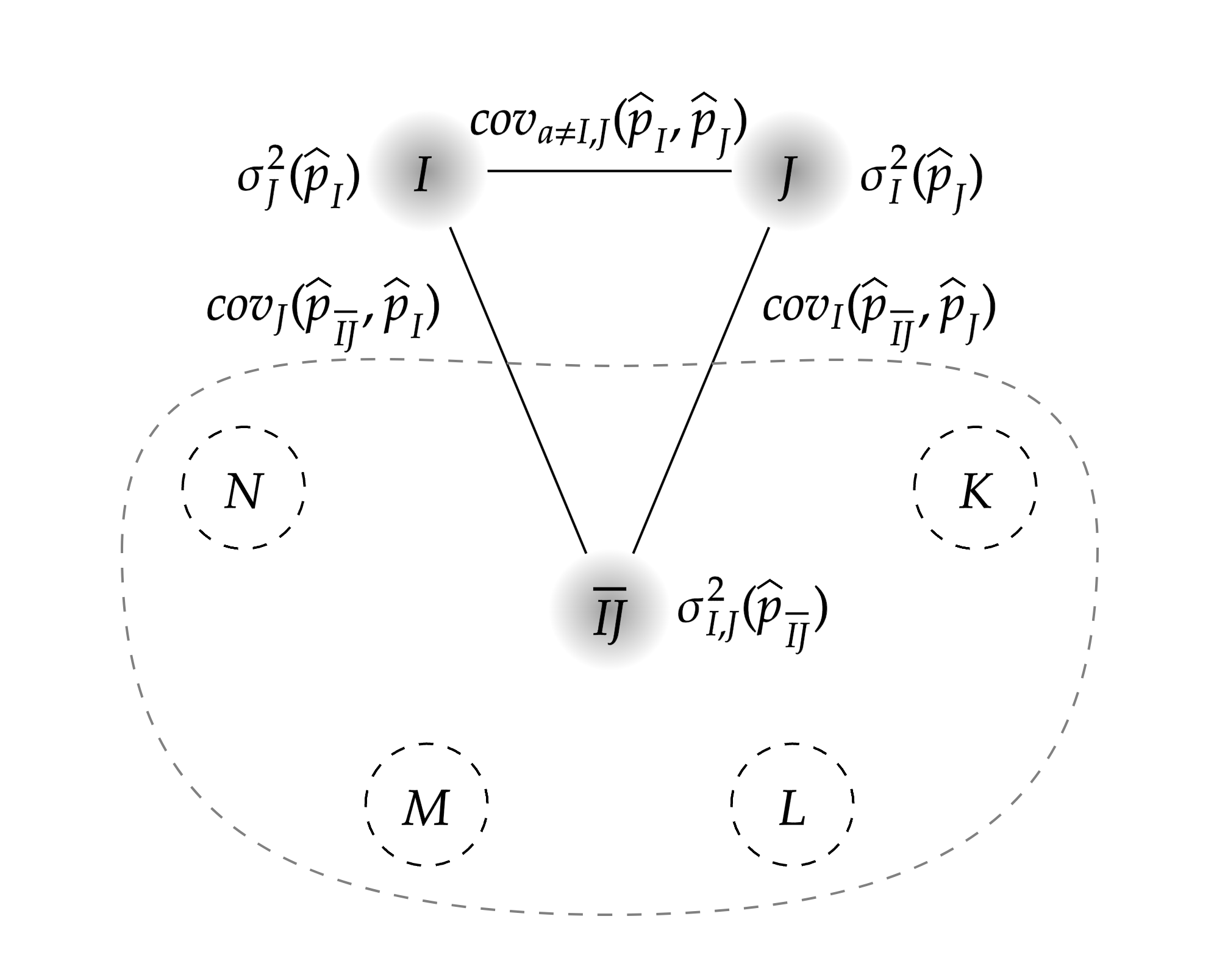}
\caption{Momentum variance of particle $J$ ($I$) and its covariance with the rest of the system, $\overline{IJ}$, from $I$'s ($J$'s) perspective, as appears in Eq.~\eqref{eq:cov_p_trnas}.}
\label{FIG4a_p_moments}
\end{subfigure}
\begin{subfigure}{6cm}
\centering
\includegraphics[width=1\linewidth]{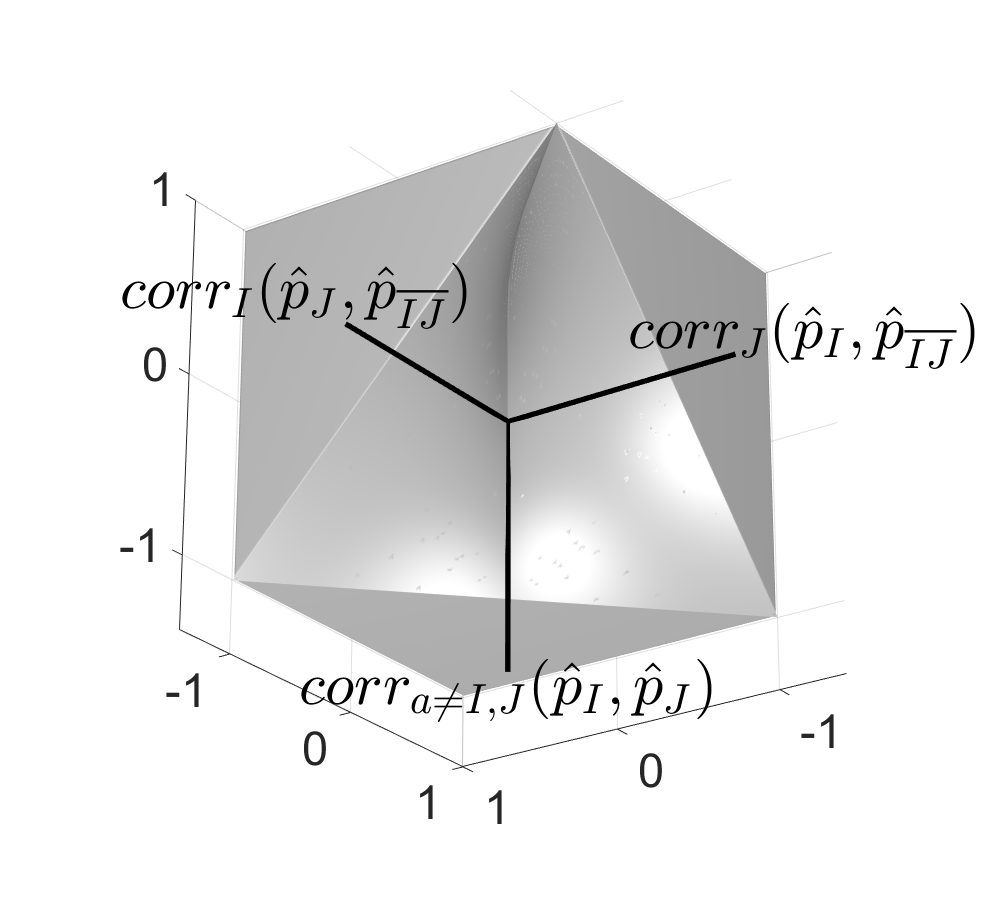}
\caption{The forbidden region (in gray) for the momentum correlations given in Eq.~\eqref{eq_p_corr_ineq}}
\label{FIG4b_p_corr_ineq_3D}
\end{subfigure}
\caption{Momentum second moments between any two QRFs, $I$ and $J$, and the rest of the system, $\overline{IJ}$.
}
\label{FIG4_momentum_moments}
\end{figure}

According to Eq.~\eqref{eq:sigma(p)_transf}, if particle $I$ has a definite momentum from any perspective, then, in the QRF associated with it, the total momentum of the described particles is also definite, $\sigma_{a\ne I}(\hat{p}_I)=\sigma_{I}(\hat{p}_{\bar I})=0$. In this case  $\sum_{a\ne I}\sigma^2_I(\hat{p}_a)=
\sum_{b\ne c \ne I}\text{cov}_I(\hat{p}_b,\hat{p}_c)$.
\\\\
According to Eq.~\eqref{eq:cov_p_trnas}, a certain particle cannot be uncorrelated with all the other particles, in all frames, and at least one of the correlations should be negative unless its momentum variance is zero.
For the case when $\sigma_{a\ne I}(\hat{p}_I)\ne0$ and $\sigma_{a\ne J}(\hat{p}_J)\ne0$, we can write the following correlations equality,
\begin{equation}\begin{aligned}
\text{corr}_{a\ne I,J}(\hat{p}_I,\hat{p}_J)=
-\text{corr}_{I}(\hat{p}_{\bar I},\hat{p}_J)=
-\text{corr}_{J}(\hat{p}_{\bar J},\hat{p}_I).
\label{eq:trans_p_corr}
\end{aligned} \end{equation}

If $J$ is not correlated in momentum with all the remaining particles, from $I$'s perspective, and vice versa, 
$
\text{cov}_I(\hat{p}_{\overline{IJ}},\hat{p}_J)=
\text{cov}_J(\hat{p}_{\overline{IJ}},\hat{p}_I)=0
$,
then, their reciprocal variance description coincide, 
$\sigma_I^2(\hat{p}_J)=\sigma_J^2(\hat{p}_I)$. In this case, they are maximally correlated in momentum, $\text{corr}_{a\ne I,J}(\hat{p}_I,\hat{p}_J)=1$, and, according to Eq.~\eqref{eq:trans_p_corr}, see each other maximally anti-correlated with the center of mass momentum of the remaining environment. Alternatively, if both, $I$ and $J$ frames, are certain about the rest of the system momentum, $\sigma_{I,J}(\hat{p}_{\overline{IJ}})=0$, then the reciprocal momentum variances coincide, $\sigma_I^2(\hat{p}_J)=\sigma_J^2(\hat{p}_I)$,  according to Eq.~\eqref{eq:var_p_ineq}.

\subsection{Perspective dependence of purity}
\label{sec:QRF_sec_purity}
In the current paper, the total state of the system is assumed to be pure from any perspective.
It is known that, in general, the subsystems of a multipartite system described by a pure state are represented by mixed states. But, do different QRFs agree about the purity of a certain subsystem from their perspectives? To address this question let us restrict the discussion in this subsection to n-mode Gaussian states where the purity is determined by the determinant of the covariance matrix of the subsystem under investigation:
\begin{equation}
    \mu(\rho_G)=\frac{1}{2^n\sqrt{\det {\bm \Sigma}}}.
\end{equation}
Using Eqs.~\eqref{app:eq_var_cov_x_1} and~\eqref{app_eq_cov(x,p)_trans} we see that the determinants of the position-momentum covariance matrices, associated with a certain particle, say $K$, from different perspectives, say $I$'s and $J$'s, do not coincide, in general,
\begin{equation}
\begin{aligned}
&\det{\bm \Sigma}_{(I)\left\{K\right\}}^{(x,p)}=\\
&\sigma_I^2(\hat{x}_K)\sigma_I^2(\hat{p}_K)-
\text{cov}_I(\hat{x}_K,\hat{p}_K)
\text{cov}_I(\hat{p}_K,\hat{x}_K)=\\
&\left[\sigma_J^2(\hat{x}_K)+
\sigma_J^2(\hat{x}_I)-2\text{cov}_J(\hat{x}_I,\hat{x}_K)
\right]
\sigma_J^2(\hat{p}_K)-\\
&\left[
\text{cov}_J(\hat{x}_K,\hat{p}_K)-
\text{cov}_J(\hat{x}_I,\hat{p}_K)
\right]\\
&\left[
\text{cov}_J(\hat{p}_K,\hat{x}_K)-\text{cov}_J(\hat{p}_K,\hat{x}_I)
\right]\ne \det{\bm \Sigma}_{(J)\left\{K\right\}}^{(x,p)}.
\label{app_eq_det(CM_1)_QRF_dep}
\end{aligned}
\end{equation}
Hence, in general, different QRFs do not agree about the purity of the same subsystem, due to correlations of the QRFs with the described object, from each others' perspectives, 
$\text{cov}_J(\hat{x}_I,\hat{x}_K)$, $\text{cov}_J(\hat{x}_I,\hat{p}_K)$,
and reciprocal position variance, $\sigma_J^2(\hat{x}_I)$.

According to Eqs.~\eqref{eq_var_cov_x}-\eqref{eq_cov_x_2} one can see that if $I$ and $J$ are approximately localized in each others' frames, i.e. $\sigma_I^2(\hat{x}_J)\approx 0$, then the spatial moments of all the other particles coincide from their perspectives,
\begin{equation}\begin{aligned}
\sigma_I^2(\hat{x}_K)=\sigma_J^2(\hat{x}_K),
\end{aligned}\end{equation} 
\begin{equation}\begin{aligned}
\text{cov}_I(\hat{x}_K,\hat{x}_L)=
\text{cov}_J(\hat{x}_K,\hat{x}_L),
\end{aligned}\end{equation} 
and, their covariances with other particles, from each other's perspectives, vanish,
\begin{equation}\begin{aligned}
\text{cov}_I(\hat{x}_J,\hat{x}_K)=
\text{cov}_J(\hat{x}_I,\hat{x}_K) \approx 0,
\end{aligned}\end{equation} 
where $K,L\in\mathfrak{N}\setminus\left\{I,J\right\}$.
Additionally, if the mixed position-momentum covariances coincide in both QRFs,
\begin{equation}
\text{cov}_J(\hat{x}_K,\hat{p}_L)=
\text{cov}_I(\hat{x}_K,\hat{p}_L),
\end{equation}
that gives, according to Eq.~\eqref{app_eq_cov(x,p)_trans},
\begin{equation}
\text{cov}_J(\hat{x}_I,\hat{p}_K)=
\text{cov}_I(\hat{x}_J,\hat{p}_K)\approx0,
\end{equation}
Under these conditions, QRFs $I$ and $J$ may be considered equivalent in describing any sub-system within a Gaussian state. Hence, they are also equivalent when addressing the corresponding purity.

We comment that the situation above indicates a sufficient but not necessary set of conditions to ensure equivalence of QRFs.
A deeper study regarding the equivalence of perspectives, e.g., the necessary and sufficient conditions under which descriptions from different QRFs coincide, lies beyond the scope of this work and will be explored in a future work.

\subsection{Invariants under changing QRFs at the initial instant}
\label{Sec_invariants}

\subsubsection{The variance-based entanglement criteria
% are invariant under QRF transformations
}
The well-known variance‐based entanglement criteria utilize EPR‐like operators, $\hat{x}_1-\hat{x}_2$ and $\hat{p}_1+\hat{p}_2$.
By analyzing the variances of these operators, one can derive criteria based on a product, introduced in~\cite{Reid1989} and a sum, introduced  in~\cite{Duan2000,Simon2000}. For Gaussian states, which are entirely described by their second moments, separability is assured if
\begin{equation}
   \text{C}^\text{prod}_{12}\coloneqq \sigma^2(\hat{x}_1-\hat{x}_2)\sigma^2(\hat{p}_1+\hat{p}_2)\ge1,
\label{eq_var_criteria_prod}
\end{equation}
or, 
\begin{equation}
    \text{C}^\text{sum}_{12}\coloneqq\sigma^2(\hat{x}_1-\hat{x}_2)+\sigma^2(\hat{p}_1+\hat{p}_2)\ge2.
\label{eq_var_criteria_sum}
\end{equation}
In other words, any violation of these inequalities is a definitive indication of entanglement. For non‑Gaussian states, however, satisfying the inequalities does not guarantee separability because important information may reside in higher-order moments. Nonetheless, their violation still provides a sufficient condition for the presence of entanglement~\cite{Reid1989,Duan2000,Simon2000}.

In the spatial QRF formalism, described in the current paper, the momentum moments coincide in all QRFs~\eqref{app:eq_f(p_I,p_J)}, at the initial instant, hence,
\begin{equation}
    \sigma_I^2(\hat{p_K}+\hat{p}_L)(0)=
    \sigma_J^2(\hat{p_K}+\hat{p}_L)(0).
\end{equation}
Additionally, the variance of the positions' difference of any two particles, is perspective-invariant, as we show in Eq.~\eqref{app:eq_var_cov_x_1},
\begin{equation}
    \sigma_I^2(\hat{x_K}-\hat{x}_L)(0)=
    \sigma_J^2(\hat{x_K}-\hat{x}_L)(0),
\end{equation}
Hence, both the product~\eqref{eq_var_criteria_prod} and the sum~\eqref{eq_var_criteria_sum} criteria are invariant under QRF transformations, at the initial instant,
\begin{equation}
    \text{C}^\text{prod/sum}_{(I)KL}(0)=
    \text{C}^\text{prod/sum}_{(J)KL}(0).
\end{equation}
This means that if any two particles are entangled according to the variance-based conditions from one perspective, it is true from any other. In the non-interacting case it holds for any $t$.

\subsubsection{The determinant of the total covariance matrix is perspective independent when the uncertainty relations are not}\label{Sec_det_cov_inv}
Following the formulation of the perspective-dependent expectation values in the previous section, we may write the relation of the position and momentum second moments, as shown in Appendix~\ref{app_sec_EV_UR}.
According to Eqs.~\eqref{eq_var_x_QRF_dep} and~\eqref{eq:var_p_inv}, we see that in general the position-momentum uncertainty relation of a certain particle, say $K$,
\begin{equation}\begin{aligned}
\sigma_I^2(\hat{x}_K)\sigma_I^2(\hat{p}_K)\ne
\sigma_J^2(\hat{x}_K)\sigma_J^2(\hat{p}_K),
\end{aligned}\end{equation} 
is not the same in different frames of reference, $I$ and $J$ in the QRF formalism, as opposed to the non-relational one in Eq.~\eqref{eq_UR_non-rel}.
What is invariant under the change of perspectives, as was shown in Appendix~\ref{app:CM_inv_under_QRF_chng}, Eqs.~\eqref{app_eq_Sigma_r_inv} and~\eqref{app_eq_det_Sigma_inv}, in the most general case, are the initial value of the determinants of the total position and momentum covariance matrices,
\begin{equation}
\det \widetilde{\bm{\Sigma}}^{(r)}_{(I)}(0)=
\det \widetilde{\bm{\Sigma}}^{(r)}_{(J)}(0),
\label{app_eq_Sigma_r_inv}
\end{equation}
where $r=x,p$, and the combined position-momentum one,
\begin{equation}
\det\bm{\Sigma}^{(x,p)}_{(I)}(0)=
\det\bm{\Sigma}^{(x,p)}_{(J)}(0).
% \equiv \mathfrak{D}_0.
\label{eq:det_cov_QRF_inv}
\end{equation}
\begin{figure}[h]
\centering
\begin{subfigure}{2.7cm}
\centering
\includegraphics[width=1\textwidth]{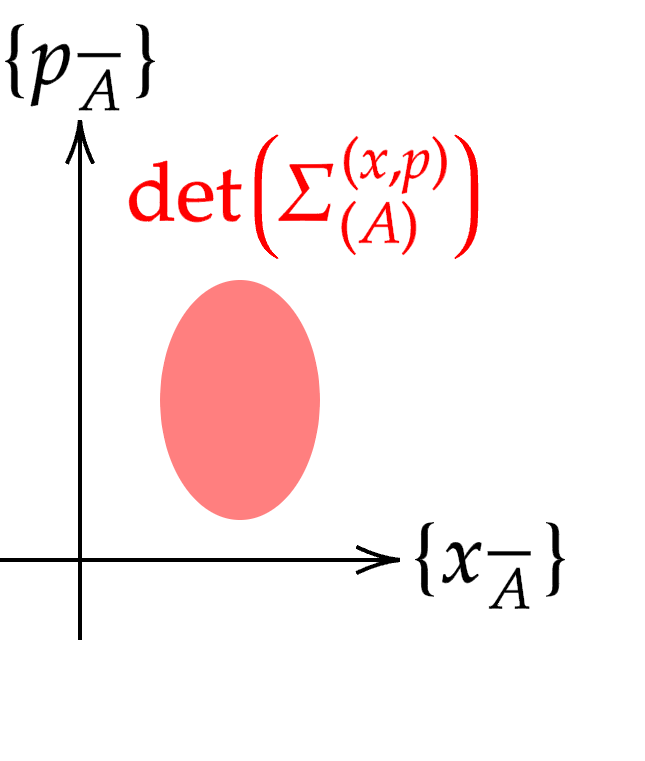}
\caption{$A$'s perspective.}
\label{FIG:det_cov_QRF_A}
\end{subfigure}
\begin{subfigure}{2.7cm}
\centering
\includegraphics[width=1\linewidth]{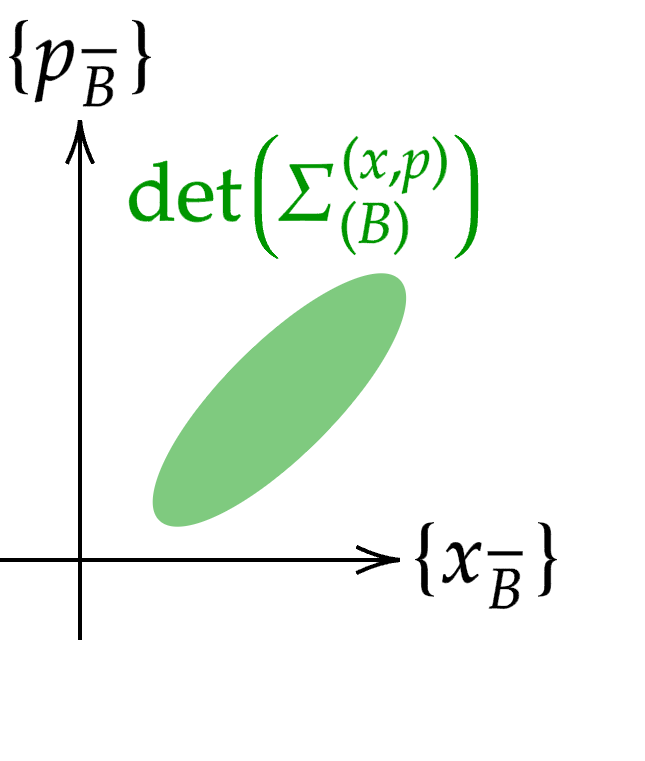}
\caption{ $B$'s perspective.}
\label{FIG:det_cov_QRF_B}
\end{subfigure}
\begin{subfigure}{2.7cm}
\centering
\includegraphics[width=1\linewidth]{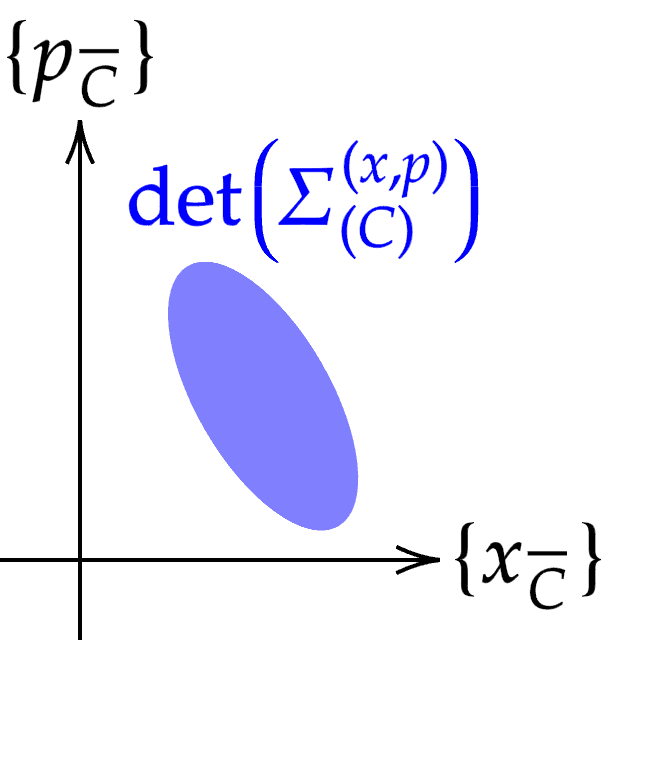}
\caption{$C$'s perspective.}
\label{FIG:det_cov_QRF_C}
\end{subfigure}
\caption{The total position-momentum uncertainty volume invariance under the change of perspectives.}
\label{FIG:det_cov_inv}
\end{figure}

Recalling that the determinant of the covariance matrix is associated with the phase space uncertainty volume, Eq.~\eqref{eq:det_cov_QRF_inv} means that this volume is invariant under the change of perspectives. 
For instance, in the case of $\mathscr{N}$ particles Gaussian state, the uncertainty volumes concerning $\mathscr{N}-1$ particles from each perspective, are of the ellipsoidal shapes, as illustrated in Fig.~\ref{FIG:det_cov_inv}.
If the particles, observed from a certain perspective, say $A$'s, 
appear in a separable state (Fig.~\ref{FIG:det_cov_QRF_A}), those observed from other perspectives, say $B$'s,  and $C$'s, might exhibit correlations (Figs.~\ref{FIG:det_cov_QRF_B} and~\ref{FIG:det_cov_QRF_C}). 
Consequently, while the shapes and orientations of the ellipsoids, associated with the total covariance matrices' determinants, may differ across perspectives, their volumes remain the same.\\

\subsection{Time evolution of the perspective-dependent position and momentum second moments}
\label{Sec_Time_evol}

\subsubsection{Position-Momentum uncertainty relations in QRFs for the case without interactions}
\label{sec:QRF_UR_without_interactions}
The reason for the deviations in moments' evolution, presented in the current section, compared to the non-relational description, is the expression of perspective-dependent velocities. 
For $\mathscr{N}$ particles system, governed by the Hamiltonian of the following form
$\hat{H}=\sum_I\frac{\hat{p}_I^2}{2m_I}+V\left(\left\{\hat{x}\right\}\right)$,
in the non-relational formalism the $I$'s particle velocity expectation value is obtained using the Heisenberg equation
\begin{equation}\begin{aligned}
\Braket{\hat{v}_J}=
\frac{d}{dt}\braket{\hat{x}_J}=
i\Braket{\left[\hat{H},\hat{x}_J\right]}=
\frac{\braket{\hat{p}_J}}{m_J},
\label{eq_NR_vel}
\end{aligned}\end{equation} 
which in another inertial Galilean frame, characterized by a classical boost velocity parameter $v_g$, becomes
\begin{equation}
\Braket{\hat{v}'_J}=
\frac{d}{dt}\Braket{\hat{x}'_J}=
\frac{\Braket{\hat{p}'_J}}{m_J}=\Braket{\hat{v}_J}+v_g.
\label{eq_nonrel_vel_trans}
\end{equation}
In the QRFs formalism, the velocity expectation value of particle $J$ from the perspective of particle $I$ is defined as follows
\begin{equation}
    \begin{aligned}
\braket{\hat{v}_{J(I)}}_I=
\frac{d}{dt}\langle\hat{x}_J\rangle_I=
&i\Braket{\left[\hat{H}_{{\bar I}},\hat{x}_J\right]}_I=
\Braket{\frac{\hat{p}_J}{m_J}+\frac{\hat{p}_{\bar I}}{m_I}}_I
% \ne \Braket{\frac{\hat{p}_J}{m_J}}_I
    \end{aligned}
\label{eq_QRF_vel}
\end{equation}
Due to particle interactions, the transformation to a different reference frame is not as straightforward as in the non-relational case. In the case without interactions, when all QRFs may be considered  inertial, using Eq.~\eqref{app_eq_vel_trans} we may write,
\begin{equation}
\left<\hat{v}_{J(K)}\right>_K=
\left<\hat{v}_{J(I)}\right>_I-\left<\hat{v}_{K(I)}\right>_I=
\left<\hat{v}_{J(I)}\right>_I+
\left<\hat{v}_{I(K)}\right>_K.
\label{eq_QRF_vel_trans}
\end{equation}
By comparison with the Galilean transformation in Eq.~\eqref{eq_nonrel_vel_trans}, we observe that the classical parameter, $v_g$, representing the velocity between different reference frames is replaced by a quantum observable.

The evolution of the position-momentum uncertainty relation,
\begin{widetext}
\begin{equation}\begin{aligned} 
&\sigma^2_I(\hat{x}_J)(t)\sigma^2_I(\hat{p}_J)(t)=
\left(\sigma^2_I(\hat{x}_J)(0)+
t\Big[
\text{cov}_I\left(\hat{x}_J,\hat{v}_{J(I)}\right)+
\text{cov}_I\left(\hat{v}_{J(I)},\hat{x}_J\right)
\Big]+
t^2\sigma^2_I\left(\hat{v}_{J(I)}\right)
\right)\sigma^2_I(\hat{p}_J)(0)=\\
&\sigma^2_I(\hat{x}_J)(0)\sigma^2_I(\hat{p}_J)(0)+\\
&t\left[
\frac{1}{m_J}
\Big(
\text{cov}_I(\hat{x}_J,\hat{p}_J)_I(0)+\text{cov}_I(\hat{p}_J,\hat{x}_J)_I(0)
\Big)
+
\frac{1}{m_I}\sum_{L\ne I}
\Big(
\text{cov}_I(\hat{x}_J,\hat{p}_L)_I(0)+
\text{cov}_I(\hat{p}_L,\hat{x}_J)_I(0)
\Big)
\right]\sigma^2_I(\hat{p}_J)(0)+
\\
&t^2\left(
\frac{1}{m_J^2}\sigma^2_I(\hat{p}_J)(0)+
2\frac{1}{m_Im_J}\sum_{L\ne I}\text{cov}_I(\hat{p}_J,\hat{p}_L)(0)+
\frac{1}{m_I^2}\sum_{K,L\ne I}\text{cov}_I(\hat{p}_K,\hat{p}_L)(0)
\right)\sigma^2_I(\hat{p}_J)(0),
\label{eq:QRF_UR}
\end{aligned}\end{equation} 
and, the determinant of the associated covariance matrix
\begin{equation}
    \begin{aligned}
&\det\Big(\bm{\Sigma}_{(I)\left\{J\right\}}^{(x,p)}(t)\Big)=
\sigma^2_I(\hat{x}_J)(t)\sigma^2_I(\hat{p}_J)(t)-
\text{cov}_I(\hat{p}_J,\hat{x}_J)(t)
\text{cov}_I(\hat{x}_J,\hat{p}_J)(t)=\\
&\sigma^2_I(\hat{x}_J)(0)\sigma^2_I(\hat{p}_J)(0)-
\text{cov}_I(\hat{p}_J,\hat{x}_J)(0)
\text{cov}_I(\hat{x}_J,\hat{p}_J)(0)\\
&t\frac{1}{m_I}\sum_{L\ne I}
\Bigg[
\sigma^2_I(\hat{p}_J)(0)\Big( \text{cov}_I(\hat{x}_J,\hat{p}_L)_I(0)+
\text{cov}_I(\hat{p}_L,\hat{x}_J)_I(0) \Big)-
\text{cov}_I(\hat{p}_J,\hat{p}_L)(0)
\Big(
\text{cov}_I(\hat{x}_J,\hat{p}_J)(0)+
\text{cov}_I(\hat{p}_J,\hat{x}_J)(0) \Big)
\Bigg]+
\\
&t^2\frac{1}{m_I^2}\sum_{K,L\ne I}
\Bigg[
\sigma^2_I(\hat{p}_J)(0)\text{cov}_I(\hat{p}_K,\hat{p}_L)(0)-
\text{cov}_I(\hat{p}_J,\hat{p}_K)(0)
\text{cov}_I(\hat{p}_J,\hat{p}_L)(0)
\Bigg],
\label{eq:QRF_CM}
    \end{aligned}
\end{equation}
\end{widetext}
are obtained using Appendix~\ref{app:section_QRF_UR_no_interactions}. The reason for the deviations in the terms above, even without changing perspectives, compared to the ones in the non-relational case appearing in Eqs.~\eqref{eq:sigma^2(x)sigma^2(p)_standardQM} and~\eqref{eq:det(cov(p,x)_standardQM} is the difference the between velocity expressions in Eqs.~\eqref{eq_NR_vel} and~\eqref{eq_QRF_vel}.
Without interactions, uncertainties, and correlations, in the QRF framework, concerning momentum only, are constant in time, as was shown in Eq.~\eqref{app:eq_sigma_p_inv} and~\eqref{app_eq:cov_p_const}.
But the ones where the position degrees of freedom are involved, are more complicated as shown in Eqs.~\eqref{app:eq_cov_I(x_J,p_K)(t)} and~\eqref{app:eq_cov_I(x_J,x_K)(t)}. We see that the expression for the position-momentum uncertainty relation in Eq.~\eqref{eq:QRF_UR}, in addition to time, depends also on relative velocities and correlations between all the subsystems, as well as the mass of the QRF.
The determinant of the perspective-dependent position-momentum covariance matrix in Eq.~\eqref{eq:QRF_CM}, depend on time, as opposed the non-relational expression in Eq.~\eqref{eq:det(cov(p,x)_standardQM}.
The QRF description coincides with the non-relational QM framework in Eqs.~\eqref{eq_UR_non-rel} and~\eqref{eq:det(cov(p,x)_standardQM} in the limit when the QRF's mass is much bigger than the mass of the described system, $m_I\gg m_J$,  which is reasonable, as discussed in~\cite{2020_change_of_perspective}.

% We see that the observed system, $I$, being separable from system $J$'s perspective, does not give the standard expression in Eq.~\eqref{eq:sigma^2(x)_standardQM}.

% which coincides with the one in the non-relational description if system $I$ is separable in the current QRF. The extra terms cancel each other out in this case regardless of the observer's mass.
% It is worthwhile mentioning that if a certain state in a certain frame is separable, it inevitably will be entangled with the environment in a different frame. Hence, the only limit for which the QRF description coincides with the non-relational one is when the observer's mass is much bigger then the ones of the observed systems.

\subsubsection{Time conservation of the determinant of the total covariance matrix}
\label{sec:det_cov_time_and_perspective_inv}
In Sec.~\ref{sec:QRF_UR_without_interactions}, we showed that the uncertainty relations in the QRF approach do not coincide with ones in the non-relational framework and that the determinant of the covariance matrix associated with an individual particle in the non-interacting case depends on time, as opposed to the non-relational approach. 
Now, we would like to concentrate on the conserved quantity in the QRF framework. The reason for the inconsistency with the non-relational QM is the conceptual claim for translation invariance on the frame neutral level -- the total momentum constraint, and the resulting correlations in the perspective-dependent description. 
In the case of quadratic Hamiltonian in a certain QRF
\begin{equation}\begin{aligned}
\hat{H}_{\bar A}=\frac{1}{2}\hat{\textbf{R}}_{\bar A}^T\textbf{G}_{\bar A}\hat{\textbf{R}}_{\bar A},
\end{aligned}\end{equation} 
where $\hat{\textbf{R}}_{\bar A}^T=
\left[\begin{matrix}
\hat{x}_B&\hat{x}_C&\ldots&\hat{p}_B&\hat{p}_C&\ldots&
\end{matrix}\right]$ and $\textbf{G}_{\bar A}$ is a symmetric $2(\mathscr{N}-1)\times2(\mathscr{N}-1)$ matrix.
Describing the entire system, from a certain perspective, all the correlations are taken into account. 
As a result, in this case, the determinant of the total covariance matrix is not only invariant under QRF transformation at the initial instant  (Eq.~\eqref{eq:det_cov_QRF_inv}), but also remains constant in time, as shown in Appendix~\ref{app_subsec:det_cov_time_ind}, in Eq.~\eqref{app:cov_QRF_time_indep}
\begin{equation}
\det \bm{\Sigma}_{(A)}^{(x,p)}(t)=
\det\bm{\Sigma}_{(A)}^{(x,p)}(0).
\label{eq:cov_QRF_time_indep}
\end{equation}

\section{Discussion and conclusions}\label{Discussion_Conclussions}

In this work, we explored the behavior of uncertainties, correlations and entanglement within the Quantum Reference Frame (QRF) formalism.
Our results indicate that while the reciprocal uncertainties in the positions of QRFs (when describing each other) are identical, the second-order moments, namely the variances and covariances, of the remaining system depend on the chosen reference frame.
Conversely, for momentum observables the situation is reversed: the QRFs agree on the second moments of the surrounding particles when the reciprocal ones differ.
We derived inequalities that constrain the possible values of these variances and correlations.
Importantly, even in the regime where the masses of the QRFs are much larger than those of the particles they describe, a limit in which one would expect the standard, non-relational picture to emerge as discussed in~\cite{2020_change_of_perspective}, the perspective-dependent deviations remain.
Exploring the perspective-dependence of subsystems' purities, in the case of Gaussian states, we formulated the conditions under which the purities coincide in different QRFs. Under these conditions, the QRFs agree about all the second moments, and hence may be considered equivalent in describing Gaussian states.  
Another notable finding is that the position-momentum variance-based entanglement criteria are invariant under QRF transformations, which is guaranteed by the translation invariance constraint. This means that the entanglement witness revealed by such conditions is perspective-independent.
We also demonstrated that transformations between different QRFs preserve the determinants of the full position and momentum covariance matrices, as well as the combined position-momentum ones.
This invariance ensures that the associated uncertainty volumes remain unchanged when switching between frames. Specifically, the volumes defined by position and momentum uncertainties, as well as the combined volume in phase space that encompasses both, remain invariant.

Finally, when considering time evolution, we demonstrated that the position-momentum uncertainty relation for an individual particle within a system of many non-interacting particles generally deviates from that predicted by the non-relational framework.
In the conventional, non-relational description, the uncertainty relation is determined solely by the particle’s own position-momentum covariance and its mass.
However, within the QRF framework, the translational invariance constraint introduces additional dependencies, the uncertainty relation also becomes a function of the mass of the particle acting as the QRF, as well as the position-momentum with, and momentum-momentum covariances among, all particles in the system.
In the limiting case, discussed in~\cite{2020_change_of_perspective}, where the QRF’s mass is significantly larger than that of the particle under consideration, the uncertainty relation converges to the conventional, non-relational expression. 
Moreover, for systems governed by quadratic Hamiltonians, our analysis reveals that the determinant of the complete position-momentum covariance matrix remains constant over time. Consequently, the corresponding uncertainty volume in phase space is invariant not only under transformations between reference frames but also during temporal evolution. Prospectively, we find it intriguing to further study this quantity and its fundamental significance.

The interactions described in Sec. \ref{sec:det_cov_time_and_perspective_inv}, concern Hamiltonians that are quadratic in the position and momentum operators. 
It would be worthwhile to investigate how broader classes of interactions, such as inverse-square potentials, affect both perspective-dependent and perspective-invariant quantities under QRF transformations. 

One of the natural generalizations of the results in the current manuscript can be achieved in future work by moving from spatial QRFs to spatiotemporal QRFs, introduced in~\cite{giacomini2021spacetime, suleymanov2023nonrelativistic_PRA}. Namely, it would be of interest to analyze correlations between degrees of freedom associated with the temporal frames of reference and the combined, spatiotemporal ones, especially in relativistic scenarios \cite{giacomini2019relativistic,giovannetti2023geometric}.
In addition, there is a clear affinity between the covariance-based mathematical framework employed in this work and the one familiar from Gaussian quantum information \cite{weedbrook2012gaussian} and quantum optics \cite{bello2021broadband}. It would be interesting to further utilize this similarity for theoretical reasons as well as for designing photonic setups which could demonstrate the relational description presented above. Possibly, the proposed formalism could be employed for sensing, computation and communication applications in correlated quantum networks of continuous degrees of freedom.

\section*{Acknowledgements}
We wish to thank V. Baumann, C. Brukner, M. de Gosson, 
M.P.E. Lock, L.D. Loveridge, L. Maccone, and I.L. Paiva for helpful discussions. 
Previous versions of this work were presented in FPQP24, TiQT24 and TMF24 with helpful feedback from their participants which helped crystalizing the main claims. We are especially thankful for P.A. H\"ohn, F. Mele, and C. Cepollaro for very useful comments.
% We also wish to thank M. Gramegna and all other organizers and participants of TMF2024 where this research was first presented and then further evolved. 
This work was supported by the European Union’s Horizon Europe research and innovation programme under grant agreement No. 101178170 and from the Israel Science Foundation under grant No. 2208/24. 

% \newpage
\onecolumngrid
\vspace{1cm}
\appendix
% \begin{center}
% % {\Huge Appendix}    
% \end{center}
\section{Uncertainties and correlations in non-relational quantum mechanics}
\label{sec:non-relational_UR}
In non-relational quantum mechanics the position-momentum variances, and hence their uncertainty relations, are invariant under Galilean transformations. For a system with $\mathscr{N}$ particles, the position and momentum uncertainties of a certain particle, say $I$, coincide in different Galilean frames of reference
\begin{equation}\begin{aligned}
\sigma^2(\hat{x}'_I)=
\sigma^2(\hat{x}_I),
\hspace{5mm}
\sigma^2(\hat{p}'_I)=
\sigma^2(\hat{p}_I).
\label{eq_UR_non-rel}
\end{aligned}\end{equation} 
just like the covariances concerning any two particles, say $I$ and $J$,
\begin{equation}\begin{aligned}
\text{cov}(\hat{x}'_I,\hat{x}'_J)=
\text{cov}(\hat{x}_I,\hat{x}_J),
\hspace{5mm}
\text{cov}(\hat{p}'_I,\hat{p}'_J)=
\text{cov}(\hat{p}_I,\hat{p}_J),
\hspace{5mm}
\text{cov}(\hat{x}'_I,\hat{p}'_J)=
\text{cov}(\hat{x}_I,\hat{p}_J),
\label{eq_cov_non-rel}
\end{aligned}\end{equation} 
where $\text{cov}(\hat{O}_1,\hat{O}_2)=\frac{1}{2}\braket{\{\hat{O}_1,\hat{O}_2\}}-\braket{\hat{O}_1}\braket{\hat{O}_2}$,  $\hat{x}_I\rightarrow \hat{x}'_I=
\hat{\Gamma}^\dagger(v,t)\hat{x}_I\hat{\Gamma}(v,t)=
\hat{x}_I+vt
$, 
$\hat{p}_I\rightarrow \hat{p}'_I=
\hat{\Gamma}^\dagger(v,t)\hat{p}_I\hat{\Gamma}(v,t)=
\hat{p}_I+vm_I
$, and the Galilean transformation is of the form,
\begin{equation}\begin{aligned}
\hat{\Gamma}=\exp\left[iv\sum_I(m_I\hat{x}_I-t\hat{p}_I)\right].
\label{eq:Gamma_Galilean_trans}
\end{aligned}\end{equation} 
The covariance matrix components, associated with the vector of operators, 
$\hat{\textbf{R}}=\left[\begin{matrix}
\hat{R}_1&\hat{R}_2&\cdots&
\end{matrix}\right]$, are of the form
\begin{equation}
\begin{aligned}
{\Sigma}_{ij}=
&\frac{1}{2}\Braket{
\hat{R}_i
\hat{R}_j+
\hat{R}_j
\hat{R}_i
}+
\Braket{\hat{R}_i}
\Braket{\hat{R}_j}=\frac{1}{2}\Braket{
\hat{R}_i
\hat{R}_j+
\left(\hat{R}_i
\hat{R}_j\right)^\dagger
}+
\Braket{\hat{R}_i}
\Braket{\hat{R}_j}=\\
&\frac{1}{2}\left(\Braket{
\hat{R}_i
\hat{R}_j}+
\Braket{\hat{R}_i
\hat{R}_j
}^*\right)+
\Braket{\hat{R}_i}
\Braket{\hat{R}_j},
\end{aligned}
\end{equation}
which may be written in a matrix form as follows
\begin{equation}
\bm{\Sigma}=
\frac{1}{2}\left(
\Braket{\hat{\textbf{R}}\hat{\textbf{R}}^T}+
\Braket{\hat{\textbf{R}}\hat{\textbf{R}}^T}^*\right)-
\Braket{\hat{\textbf{R}}}
\Braket{\hat{\textbf{R}}^T}.
\end{equation}

% The covariance matrix elements, associated with $\mathscr{N}$ operators, 
% $\hat{\textbf{R}}^T=
% \left[\begin{matrix}
% \hat{R}_1&\cdots&\hat{R}_\mathscr{N}
% \end{matrix}\right]$, 
% are defined as follows
% \begin{equation}    \Sigma_{ij}=
% \frac{1}{2}
% \Braket{\left\{\hat{R}_i,\hat{R}_j\right\}}-
% \Braket{\hat{R}_i}
% \Braket{\hat{R}_j}
% \end{equation}
% 
% 
% 
% The uncertainty volume in the phase,
% associated with each $\textbf{c}_{IJ}^{(x,p)}$, is also the same in all Galilean frames
% \begin{equation}\begin{aligned}
% \det\left(\textbf{c}_{IJ}^{(x,p)}\right)=
% \det\left(\textbf{c}_{IJ}^{(x',p')}\right)
% \end{aligned}\end{equation} 
Combined with the commutation relations matrix, $\left[\hat{R}_i,\hat{R}_j\right]=i\Omega_{ij}$, this gives the following positive semi-definite matrix
\begin{equation}
        \bm{\Sigma}+\frac{i}{2}\bm{\Omega}\succeq 0,
\end{equation}
which is called the Schr\"odinger-Robertson uncertainty relation. \\
Considering $\mathscr{N}$ particles system, $\mathfrak{N}\coloneqq\left\{A,B,...,\mathscr{N}\right\}$, in one dimension, the associated position and momentum operators may be written in a vector form $\hat{\textbf{R}}^T=
\left[\begin{matrix}
\hat{x}_1&\hat{p}_1&\cdots&\hat{x}_\mathscr{N}&\hat{p}_\mathscr{N}
\end{matrix}\right]$.
The associated $2\mathscr{N}\times 2\mathscr{N}$ covariance matrix,
\begin{equation}
    \begin{aligned}
\bm{\Sigma}_{\mathfrak{N}}^{(x,p)}=
\left[\begin{matrix}
\vspace{2mm}
\bm{\Sigma}_{\{A\}}^{(x,p)}&
\bm{\Gamma}_{AB}^{(x,p)}&
\bm{\Gamma}_{AC}^{(x,p)}&\cdots\\ 
\vspace{2mm}
\bm{\Gamma}_{BA}^{(x,p)}&
\bm{\Sigma}_{\{B\}}^{(x,p)}&
\bm{\Gamma}_{BC}^{(x,p)}&\cdots\\
\bm{\Gamma}_{CA}^{(x,p)}&
\bm{\Gamma}_{CB}^{(x,p)}&
\bm{\Sigma}_{\{C\}}^{(x,p)}&\cdots\\
\vdots&\vdots&\vdots&\ddots
\end{matrix}\right],
\label{eq:total_CM_NR}
\end{aligned}
\end{equation} 
where,
\begin{equation}
    \begin{aligned}
\bm{\Sigma}_{\{J\}}^{(x,p)}=
\left[\begin{matrix}
\sigma^2(\hat{x}_J)&\text{cov}(\hat{x}_J,\hat{p}_J)\\
\text{cov}(\hat{p}_J,\hat{x}_J)&\sigma^2(\hat{p}_J)
\end{matrix}\right],
\label{eq_CM(A)J}
    \end{aligned}
\end{equation}
is associated with an individual particle $J$, and,
\begin{equation}
\bm{\Gamma}_{JK}^{(x,p)}\equiv 
\left[\begin{matrix}
\text{cov}(\hat{x}_J,\hat{x}_K)&
\text{cov}(\hat{x}_J,\hat{p}_K)\\
\text{cov}(\hat{p}_J,\hat{x}_K)&
\text{cov}(\hat{p}_J,\hat{p}_K)
\end{matrix}\right],
\end{equation}
reflects the covariances between $J$ and $K$, 
is also frame invariant under the Galilean transformation given in Eq.~\eqref{eq:Gamma_Galilean_trans}
\begin{equation}\begin{aligned}
% \textbf{cov}
\bm{\Sigma}_\mathfrak{N}^{(x,p)}=\bm{\Sigma}_\mathfrak{N}^{(x',p')}.
\end{aligned}\end{equation}
According to Williamson's theorem, any covariance matrix can be diagonalized via a symplectic transformation $\textbf{S}$, 
$\bm{\Sigma}_\mathfrak{N}^{(x,p)}=\textbf{S}\left(\bigoplus_{k=1}^\mathscr{N}\nu_k\mathbb{I}_2\right)\textbf{S}^T$, 
where $\nu_k\ge0$ are the symplectic eigenvalues. Using that 
$\textbf{S}\bm{\Omega}\textbf{S}^T=\bm{\Omega}=\bigoplus_{i=1}^\mathscr{N}
\left[\begin{matrix}
    0 & 1\\-1&0
\end{matrix}\right]$, we have $\nu_k\ge\frac{1}{2}$, and hence
\begin{equation}
\det\bm{\Sigma}_\mathfrak{N}^{(x,p)}\ge\left(\frac{1}{2}\right)^{2\mathscr{N}},
\end{equation}
reflecting the minimal uncertainty volume in the phase space.\\
Focusing on an individual particle's position-momentum covariance matrix 
% \be
% \textbf{c}_{II}^{(x,p)}\equiv 
% \left[\begin{matrix}
% \text{cov}(\hat{x}_I,\hat{x}_I)&
% \text{cov}(\hat{x}_I,\hat{p}_I)\\
% \text{cov}(\hat{p}_I,\hat{x}_I)&
% \text{cov}(\hat{p}_I,\hat{p}_I)
% \end{matrix}\right]\succeq 0
% \end{aligned}\end{equation} 
we have
\begin{equation}\begin{aligned}
\det\bm{\Sigma}_{\{I\}}^{(x,p)}=
\sigma^2(\hat{x}_I)\sigma^2(\hat{p}_I)-
\text{cov}(\hat{x}_I,\hat{p}_I)
\text{cov}(\hat{p}_I,\hat{x}_I)
\ge \frac{1}{4}.
\label{eq:CM_non-relational}
\end{aligned}\end{equation} 
In the following, in Sec.~\ref{Sec_det_cov_inv}, we formulate a similar expression in a relational manner stressing the differences with the non-relational one above. \\\\
% We see above that correlations with the environment increase the uncertainty relation.\\\\
For the non-interacting case, in a certain frame, the time dependence of the position-momentum covariance
\begin{equation}\begin{aligned}
\text{cov}(\hat{x}_I,\hat{p}_I)(t)=\text{cov}(\hat{x}_I,\hat{p}_I)(0)+
\frac{t}{m_I}\sigma^2(\hat{p}_I)(0),
\label{eq:cov(p,x)_standardQM}
\end{aligned}\end{equation}  
and, uncertainty relation
\begin{equation}
    \begin{aligned}
\sigma^2(\hat{x}_I)(t)\sigma^2(\hat{p}_I)(t)=
\sigma^2(\hat{x}_I)(0)\sigma^2(\hat{p}_I)(0)+
t\frac{1}{m_I}\Big[\text{cov}(\hat{x}_I,\hat{p}_I)(0)+\text{cov}(\hat{p}_I,\hat{x}_I)(0)\Big]\sigma^2(\hat{p}_I)(0)+
t^2\frac{1}{m_I^2}\sigma^4(\hat{p}_I)(0),
    \end{aligned}
    \label{eq:sigma^2(x)sigma^2(p)_standardQM}
\end{equation}
can be obtained using the expectation values' calculations
\begin{equation}\begin{aligned}
\Braket{\hat{x}_I^a\hat{p}_J^b}(t)=
\Bra{\psi(0)}
e^{it\sum_K\frac{\hat{p}_K^2}{2m_K}}
\hat{x}_I^a\hat{p}_J^b
e^{-it\sum_K\frac{\hat{p}_K^2}{2m_K}}
\ket{\psi(0)}=
\Bra{\psi(0)}
\left(\hat{x}_I+t\frac{\hat{p}_I}{m_I}\right)^a\hat{p}_J^b
\ket{\psi(0)}.
\end{aligned}\end{equation} 
In this case, the determinant of the covariance matrix associated with a certain particle is constant in time
\begin{equation}\begin{aligned}
\det\bm{\Sigma}_I^{(x,p)}(t)&=
\sigma(\hat{x}_I)^2(t)\sigma(\hat{p}_I)^2(t)-
\text{cov}(\hat{x}_I,\hat{p}_I)(t)\text{cov}(\hat{p}_I,\hat{x}_I)(t)=\\
&=
\sigma^2(\hat{x}_I)(0)\sigma^2(\hat{p}_I)(0)-
\text{cov}(\hat{x}_I,\hat{p}_I)(0)\text{cov}(\hat{p}_I,\hat{x}_I)(0).
\label{eq:det(cov(p,x)_standardQM}
\end{aligned}\end{equation} 
In the above, we presented the description in the framework of non-relational quantum mechanics emphasizing the frame independence of the position and momentum second moments in Eqs.~\eqref{eq_UR_non-rel} and~\eqref{eq_cov_non-rel}, and their time dependence in the non-interacting case in Eqs.~\eqref{eq:cov(p,x)_standardQM},~\eqref{eq:sigma^2(x)sigma^2(p)_standardQM} and~\eqref{eq:det(cov(p,x)_standardQM}.
However, one must remember that frames of reference in such a framework are abstract entities with no physical properties, dynamics, or internal structure. Such frames cannot have interactions or correlations with the described objects, and we can define as many frames as we want.
In the relational approach, where only relative description has a meaning, i.e. description of objects from the perspective of another object, the situation is different, as we show in Sec.~\ref{sec:QRF_sec_mom}.

\section{Covariance matrices in the QRF formalism}
In $\mathscr{N}$ particles system, the position-momentum covariance matrix, from a certain perspective, say $A$'s, of $\mathscr{M}$ particles,  $\mathfrak{M}\coloneqq \left\{B,C,D,...,\mathscr{M}\right\}$, constructed using a vector of operators of the form,
$\hat{\textbf{R}}_{\bar A}=
\left[ \begin{matrix} 
\hat{x}_B&\hat{p}_B&\hat{x}_C &\hat{p}_C&\ldots&\hat{x}_\mathscr{M} &\hat{p}_\mathscr{M}
\end{matrix}\right]$,
is a symmetric $2\mathscr{M}\times2\mathscr{M}$ matrix of the form,
\begin{equation}
    \begin{aligned}
\bm{\Sigma}_{(A)\mathfrak{M}}^{(p,x)}=
\left[\begin{matrix}
\vspace{2mm}
\bm{\Sigma}_{(A)\{B\}}^{(x,p)}&
\bm{\Gamma}_{(A)BC}^{(x,p)}&
\bm{\Gamma}_{(A)BD}^{(x,p)}&\cdots\\
\vspace{2mm}
\bm{\Gamma}_{(A)CB}^{(x,p)}&
\bm{\Sigma}_{(A)\{C\}}^{(x,p)}&
\bm{\Gamma}_{(A)CD}^{(x,p)}&\cdots\\
\vspace{2mm}
\bm{\Gamma}_{(A)DB}^{(x,p)}&
\bm{\Gamma}_{(A)DC}^{(x,p)}&
\bm{\Sigma}_{(A)\{D\}}^{(x,p)}&\cdots\\
\vdots&\vdots&\vdots&\ddots
\end{matrix}\right],
\label{eq:cov_mat_A}
\end{aligned}\end{equation} 
% where
% \begin{equation}
%     \begin{aligned}
% \bm{\Sigma}_{(A)J}^{(x,p)}=
% \left[\begin{matrix}
% \sigma_A^2(\hat{x}_J)&\text{cov}_A(\hat{x}_J,\hat{p}_J)\\
% \text{cov}_A(\hat{p}_J,\hat{x}_J)&\sigma_A^2(\hat{p}_J)
% \end{matrix}\right]
% \label{eq_CM(A)J}
%     \end{aligned}
% \end{equation}
% is associated with an individual particle $J$, and
% \begin{equation}
% \bm{\Gamma}_{(A)JK}^{(x,p)}\equiv 
% \left[\begin{matrix}
% \text{cov}_A(\hat{x}_J,\hat{x}_K)&
% \text{cov}_A(\hat{x}_J,\hat{p}_K)\\
% \text{cov}_A(\hat{p}_J,\hat{x}_K)&
% \text{cov}_A(\hat{p}_J,\hat{p}_K)
% \end{matrix}\right]
% \end{equation}
% reflects the covariances between $J$ and $K$. 
% The covariance matrix in Eq.~\eqref{eq:cov_mat_A} 
which may be divided as follows
\begin{equation}
    \begin{aligned}
\bm{\Sigma}_{(A)\mathfrak{M}}^{(p,x)}=
\left[\begin{matrix}
\vspace{2mm}
\bm{\Sigma}_{(A)\{B\}}^{(x,p)}&
\textbf{C}_{(A)B}^{(x,p)}\\
\vspace{2mm}
\textbf{C}_{(A)B}^{(x,p)\dagger}&
\bm{\Sigma}_{(A)\mathfrak{M}\setminus \{B\}}^{(x,p)}
\end{matrix}\right],
\label{eq:cov_mat_A_div}
\end{aligned}\end{equation} 
where, 
\begin{equation}
    \begin{aligned}
\bm{\Sigma}_{(A)\{B\}}^{(x,p)}=
\left[\begin{matrix}
\sigma^2_A(\hat{x}_B)&\text{cov}_A(\hat{x}_B,\hat{p}_B)\\
\text{cov}_A(\hat{p}_B,\hat{x}_B)&\sigma^2_A(\hat{p}_B)
\end{matrix}\right],
    \end{aligned}
\end{equation}
represents the covariance matrix associated with particle $B$, 
% as defined in Eq.~\eqref{eq_CM(A)J}, 
$\textbf{C}_{(A)B}^{(x,p)}=\left[\begin{matrix}
\bm{\Gamma}_{(A)BC}^{(x,p)}&
\bm{\Gamma}_{(A)BD}^{(x,p)}&\cdots&
\end{matrix}\right]$
represents the covariances between particle $B$ with all the others, and, 
\begin{equation}
\bm{\Sigma}_{(A)\mathfrak{M}\setminus \{B\}}^{(x,p)}=
\left[\begin{matrix}
\vspace{2mm}
\bm{\Sigma}_{(A)\{C\}}^{(x,p)}&
\bm{\Gamma}_{(A)CD}^{(x,p)}&\cdots\\
\vspace{2mm}
\bm{\Gamma}_{(A)DC}^{(x,p)}&
\bm{\Sigma}_{(A)\{D\}}^{(x,p)}&\cdots\\
\vdots&\vdots&\ddots
\end{matrix}\right],
\end{equation}
is then a $2(\mathscr{M}-1)\times2(\mathscr{M}-1)$ covariance matrix of the rest of the system. In case when all the particles in a certain QRF, say $A$'s, are considered, $\mathfrak{M}=\mathfrak{N}\setminus \{A\}=\mathfrak{A}$, we adopt the shorthand notation, 
$\bm{\Sigma}_{(A)}\equiv\bm{\Sigma}_{(A)\mathfrak{A}}$ and 
$\bm{\Sigma}_{(A)\setminus \{B\}}\equiv\bm{\Sigma}_{(A)\mathfrak{A}\setminus \{B\}}$.
We may alternatively write a different, more useful in some cases, form of the position-momentum covariance matrix, constructed using a vector of operators of the form 
$\hat{\widetilde{\textbf{R}}}^T_{\bar A}=
\left[ \begin{matrix} 
\hat{\textbf{X}}^T_{\bar A}&\hat{\textbf{P}}^T_{\bar A}
\end{matrix}\right]$, where 
$\hat{\textbf{X}}^T_{\bar A}=
\left[ \begin{matrix} 
\hat{x}_B&\hat{x}_C&\ldots&\hat{x}_\mathscr{M}
\end{matrix}\right]$, and 
$\hat{\textbf{P}}^T_{\bar A}=
\left[ \begin{matrix} 
\hat{p}_B&\hat{p}_C&\ldots&\hat{p}_\mathscr{M}
\end{matrix}\right]$, 
\begin{equation}
\widetilde{\bm\Sigma}_{(A)\mathfrak{M}}^{(x,p)}=
\left[\begin{matrix}
\vspace{2mm}
\widetilde{\bm{\Sigma}}_{(A)\mathfrak{M}}^{(x)}&
\widetilde{\bm{\Gamma}}_{(A)\mathfrak{M}}^{(x,p)}\\
\widetilde{\bm{\Gamma}}_{(A)\mathfrak{M}}^{(x,p)}&
\widetilde{\bm{\Sigma}}_{(A)\mathfrak{M}}^{(p)}
\end{matrix}\right],
\label{eq_Sigma_tilde}
\end{equation}
where the position and momentum $\mathscr{M}\times\mathscr{M}$ covariance matrices are separate,
\begin{equation}
\widetilde{\bm{\Sigma}}_{(A)\mathfrak{M}}^{(r)}=
\left[\begin{matrix}
\sigma^2_A(\hat{r}_B)&\text{cov}_A(\hat{r}_{B},\hat{r}_C)&\cdots\\
\text{cov}_A(\hat{r}_{C},\hat{r}_B)&
\sigma^2_A(\hat{r}_C)&\cdots\\
\vdots&\vdots&\ddots
\end{matrix}\right],
\label{app_eq_Sigma_r_tilde}
\end{equation}
with $r=x,p$, and the off-diagonal ones 
\begin{equation}
\widetilde{\bm{\Gamma}}_{(A)\mathfrak{M}}^{(x,p)}=
\left[\begin{matrix}
\text{cov}_A(\hat{x}_B,\hat{p}_B)&
\text{cov}_A(\hat{x}_B,\hat{p}_C)&\cdots\\
\text{cov}_A(\hat{x}_C,\hat{p}_B)&
\text{cov}_A(\hat{x}_C,\hat{p}_C)&\cdots\\
\vdots&\vdots&\ddots
\end{matrix}\right],
\end{equation}
contain the mixed position-momentum terms.
The position and momentum covariance matrices in Eq.~\eqref{app_eq_Sigma_r_tilde} can also be divided as in Eq.~\eqref{eq:cov_mat_A_div},
\begin{equation}
    \begin{aligned}
\widetilde{\bm{\Sigma}}_{(A)\mathfrak{M}}^{(r)}=
\left[\begin{matrix}
\vspace{2mm}
\sigma_A^2(\hat{r}_B)&
\widetilde{\textbf{C}}_{(A)B}^{(r)}\\
\vspace{2mm}
\widetilde{\textbf{C}}_{(A)B}^{(r)\dagger}&
\widetilde{\bm{\Sigma}}_{(A)\setminus \{B\}}^{(r)}
\end{matrix}\right]
\label{app_eq_cov_mat_tilde_A_div},
\end{aligned}\end{equation} 
where, 
\begin{equation}
\widetilde{\textbf{C}}_{(A)B}^{(r)}=\left[\begin{matrix}
\text{cov}_A(\hat{r}_B,\hat{r}_C)&\text{cov}_A(\hat{r}_B,\hat{r}_D)&\ldots&
\end{matrix}\right],
\label{app_eq_C_r}
\end{equation}
and
\begin{equation}
\widetilde{\bm{\Sigma}}_{(A)\setminus \{B\}}^{(r)}=
\left[\begin{matrix}
\sigma^2_A(\hat{r}_C)&\text{cov}_A(\hat{r}_{C},\hat{r}_D)&\cdots\\
\text{cov}_A(\hat{r}_{D},\hat{r}_C)&
\sigma^2_A(\hat{r}_D)&\cdots\\
\vdots&\vdots&\ddots
\end{matrix}\right].
\label{app_eq_Sigma_r}
\end{equation}
Both, mixed and separated forms have coinciding determinants, 
$\det\left(\widetilde{\bm\Sigma}_{(A)\mathfrak{M}}^{(x,p)}\right)=
\det\left(\bm{\Sigma}_{(A)\mathfrak{M}}^{(x,p)}\right)$.
\section{Position and momentum first and second moments in the QRFs framework at the initial instant}
\label{app_sec_EV_UR}
% We now explore the perspective-dependent expressions describing expectation values and variances of the relevant operators in the framework just introduced. Different from the simplification considered in the previous section, we will assume a general number of subsystems, i.e., $\mathfrak{N}\coloneqq\{A,B,C,\hdots\}$.

% Recall that the expression for the expectation value of an operator $\hat O_{\bar I}$ (that does not depend explicitly on time) from the perspective of the system $I$ is
% % \begin{equation}
% %     \braket{\hat{O}_{\bar I}}_I(t_I) = 
% %     \bra{\psi_{\bar I}(t_I)}
% %     \hat{O}_{\bar I}
% %     \ket{\psi_{\bar I}(t_I)}.
% % \end{equation}
% \begin{equation}
% \langle\hat{O}\rangle_J(t)=
% \langle\psi_{\bar J}(t)|\hat{O}|\psi_{\bar J}(t)\rangle=
% \langle\psi_{\bar J}(t=0) e^{it\hat{H}_{\bar J}}
% |\hat{O}|
% e^{-it\hat{H}_{\bar J}} \ket{\psi_{\bar J}(t=0)}
% \end{equation}

% Similarly, its variance is
% \begin{equation}
%     \sigma^2(\hat{O}_{\bar I})_{I}(t)=
%     \braket{\hat{O}_{\bar I}^2}_I(t) -
%     \braket{\hat{O}_{\bar I}}_I^2(t).
% \end{equation}
% When it is sufficiently clear, we omit the dependency on time.

% Moreover, for any integrable function $f(P)$ and for every $I,J\in\mathfrak{N}$, it holds that
% \begin{equation}
%     \label{eq:identity-any-frame}
%     \begin{aligned}
%     \int  dP &\delta(P_T) f(P) 
%     =\int dP_{\bar I} f(p_I=-p_{\bar I},P_{\bar I}) 
%     =\int dP_{\bar I} f_{\bar I}(P_{\bar I})=\int dP_{\bar J} f_{\bar J}(P_{\bar J})
%     \end{aligned}
% \end{equation}

\subsection{Position}
\label{app_subsec_position}
The expectation value of an arbitrary function of the position of particle, say $K$, in the reference frame of particle, say $I$, at the initial instant is of the form,
\begin{equation}\begin{aligned}
\braket{f(\hat x_K)}_I(0)=
&\langle\psi_{\bar I}(0)|f(\hat x_K)|\psi_{\bar I}(0)\rangle=
\int dP_{\bar I}
\psi_{\bar I}^*(P_{\bar I})
f\left(i\frac{d}{dp_K}\right) \psi_{\bar I}(P_{\bar I})=
\int dP_{\bar I}
\left[ \psi^*(P)
f\left(i\frac{d}{dp_K}-i\frac{d}{dp_I}\right) \psi(P)\right]_{\bar I}=\\
-&\int dP_{\bar K}
\left[ \psi^*(P)
f\left(i\frac{d}{dp_I}-i\frac{d}{dp_K}\right) \psi(P)\right]_{\bar K}=\\
&\int dP_{\bar J}
\left[ \psi^*(P)
f\left(i\frac{d}{dp_K}-i\frac{d}{dp_J}+i\frac{d}{dp_J}-i\frac{d}{dp_I}\right) \psi(P)\right]_{\bar J},
\label{eq:<f(x_I)>_J(0)}
\end{aligned}\end{equation} 
where we used the derivative of a constrained function,
\begin{equation}
f\left(i\frac{d}{dp_K}\right)\psi_{\bar I}(P_{\bar I})=
f\left(i\frac{d}{dp_K}\right)\psi\left(p_A,...,\,\,p_I=-\sum_{i\ne I} p_i\,\,,...,p_K,...\right)=
\left[
f\left(i\frac{d}{dp_K}-i\frac{d}{dp_I}\right)\psi(P)
\right]_{\bar I},
\end{equation}
and the Eq.~\eqref{eq:identity-any-frame} for changing the frame of the integral. Hence, Eq.~\eqref{eq:<f(x_I)>_J(0)} gives the following QRF transformations
\begin{equation}\begin{aligned}
\Braket{f(\hat{x}_J)}_I(0)=
\Braket{f(-\hat{x}_I)}_J(0)=
\Braket{f(\hat{x}_J-\hat{x}_I)}_K(0).
\label{app:eq_trans_f(x)}
\end{aligned}\end{equation} 
That allows us to write the transformations for the position expectation values
\begin{equation}\begin{aligned}
\braket{\hat{x}_J}_I(0)=-\braket{\hat{x}_I}_J(0)=
\braket{\hat{x}_J}_K(0)-\braket{\hat{x}_I}_K(0),
\end{aligned}\end{equation} 
and, those of the second moments, the variance,
\begin{equation}\begin{aligned}
\sigma^2_I(\hat{x}_J)(0)=\sigma^2_J(\hat{x}_I)(0)=
\sigma^2_K(\hat{x}_I-\hat{x}_J)(0)=
\sigma^2_K(\hat{x}_I)(0)+\sigma^2_K(\hat{x}_J)(0)-
2\text{cov}_K(\hat{x}_I,\hat{x}_J)(0),
\label{app:eq_var_cov_x_1}
\end{aligned}\end{equation} 
% which gives the following inequality bound for the position-variance
% \begin{equation}\begin{aligned}
% \left(\sigma_K(\hat{x}_I)-\sigma_K(\hat{x}_J)\right)^2
% \le
% \sigma_I^2(\hat{x}_J)
% \le 
% \left(\sigma_K(\hat{x}_I)+\sigma_K(\hat{x}_J)\right)^2
% \label{app:eq_var_x_ineq}
% \end{aligned}\end{equation} 
and the covariances,
\begin{equation}
\begin{aligned}
\text{cov}_I(\hat{x}_J,\hat{x}_K)(0)=
&\text{cov}_J(-\hat{x}_I,\hat{x}_K-\hat{x}_I)(0)=
-\text{cov}_J(\hat{x}_I,\hat{x}_K)(0)+
\sigma^2_J(\hat{x}_I)(0)=\\
&\text{cov}_K(\hat{x}_J-\hat{x}_I,-\hat{x}_I)(0)=
-\text{cov}_K(\hat{x}_J,\hat{x}_I)(0)+
\sigma^2_K(\hat{x}_I)(0),
\end{aligned}
\label{app:eq_cov_x_1}
\end{equation}
and,
\begin{equation}
\begin{aligned}
\text{cov}_I(\hat{x}_J,\hat{x}_K)(0)=
&\text{cov}_L(\hat{x}_J-\hat{x}_I,\hat{x}_K-\hat{x}_I)(0)=\\
&\text{cov}_L(\hat{x}_J,\hat{x}_K)(0)-
\text{cov}_L(\hat{x}_J,\hat{x}_I)(0)-
\text{cov}_L(\hat{x}_K,\hat{x}_I)(0)+
\sigma^2_L(\hat{x}_I)(0).
\end{aligned}
\label{app:eq_cov_x_2}
\end{equation}
{By substituting Eq.~\eqref{app:eq_cov_x_1} into Eq.~\eqref{app:eq_cov_x_2} we have,
\begin{equation}
\text{cov}_I(\hat{x}_J,\hat{x}_K)(0)=
-\text{cov}_J(\hat{x}_L,\hat{x}_K)(0)
+\text{cov}_J(\hat{x}_L,\hat{x}_I)(0)
+\text{cov}_I(\hat{x}_K,\hat{x}_L)(0).
\label{app:eq_cov_x_3}
\end{equation}
The mixed position-momentum covariance transforms as follows
\begin{equation}
\text{cov}_I(\hat{x}_K,\hat{p}_L)=
\text{cov}_J(\hat{x}_K,\hat{p}_L)-
\text{cov}_J(\hat{x}_I,\hat{p}_L).
\label{app_eq_cov(x,p)_trans}
\end{equation}
}
% \begin{equation}
% \begin{aligned}
% \text{cov}_J(\hat{x}_I,\hat{x}_K)(t)=
% &\text{cov}_I(-\hat{x}_J,\hat{x}_K-\hat{x}_J)(t)=
% -\text{cov}_I(\hat{x}_J,\hat{x}_K)(t)+
% \sigma^2_I(\hat{x}_J)(t)=\\
% &\text{cov}_K(\hat{x}_I-\hat{x}_J,-\hat{x}_J)(t)=
% -\text{cov}_K(\hat{x}_I,\hat{x}_J)(t)+
% \sigma^2_K(\hat{x}_J)(t)
% \end{aligned}
% \end{equation}
Using the above, first, let us explore the position covariances when switching to the reference frame of one of the described particles in Eq.~\eqref{app:eq_cov_x_1}. 
In terms of variances and correlations, for any three particles, we have,
\begin{subequations}
    \begin{align}
&\sigma_1-\sigma_2 c_3-\sigma_3c_2=0\label{app:eq_var_corr_x_short_a},\\
&\sigma_2-\sigma_1 c_3-\sigma_3c_1=0\label{app:eq_var_corr_x_short_b},\\
&\sigma_3-\sigma_2 c_1-\sigma_1c_2=0\label{app:eq_var_corr_x_short_c},
    \end{align}
\label{app:eq_var_corr_x_short}
\end{subequations}
% \begin{equation}\begin{aligned}
% \sigma_I(\hat{x}_J)-
% \sigma_J(\hat{x}_K)\text{corr}_J(\hat{x}_I,\hat{x}_K)-
% \sigma_K(\hat{x}_I)\text{corr}_I(\hat{x}_J,\hat{x}_K)=0
% \end{aligned}\end{equation} 
% \begin{equation}\begin{aligned}
% \sigma_J(\hat{x}_K)-
% \sigma_I(\hat{x}_J)\text{corr}_J(\hat{x}_I,\hat{x}_K)-
% \sigma_K(\hat{x}_I)\text{corr}_K(\hat{x}_I,\hat{x}_J)=0
% \end{aligned}\end{equation}
% \begin{equation}\begin{aligned}
% \sigma_K(\hat{x}_I)-
% \sigma_J(\hat{x}_K)\text{corr}_K(\hat{x}_I,\hat{x}_J)-
% \sigma_I(\hat{x}_J)\text{corr}_I(\hat{x}_J,\hat{x}_K)=0
% \end{aligned}\end{equation} 
where, 
$\sigma_1\equiv\sigma_I(\hat{x}_J)$,
$\sigma_2\equiv\sigma_J(\hat{x}_K)$,
$\sigma_3\equiv\sigma_K(\hat{x}_I)$,
$c_1\equiv\text{corr}_K(\hat{x}_I,\hat{x}_J)$,
$c_2\equiv\text{corr}_I(\hat{x}_J,\hat{x}_K)$,
$c_3\equiv\text{corr}_J(\hat{x}_K,\hat{x}_I)$. 
% we can immediately say that 
% \begin{equation}
% \sigma_i\le \sigma_j+\sigma_k
% \end{equation}
First, that gives the triangle inequality for the variances,
\begin{equation}
|\sigma_i-\sigma_j|\le \sigma_k\le \sigma_i+\sigma_j
\end{equation}
where $i,j,k=1,2,3$. Using Eqs.
~\eqref{app:eq_var_corr_x_short_a},
~\eqref{app:eq_var_corr_x_short_b} and
~\eqref{app:eq_var_corr_x_short_c} we have, 
\begin{equation}
(\sigma_i+\sigma_j)(1-c_k)=\sigma_k(c_i+c_j),
\end{equation}
and
\begin{equation}
\sigma_i(1-c_ic_j)=\sigma_k(c_ic_j+c_k),
\end{equation}
which gives the following inequalities for the correlations,
\begin{equation}
c_i+c_j\ge0,
\label{app:eq_corr_ineq_2}
\end{equation}
and,
\begin{equation}
c_ic_j+c_k\ge0.
\label{app:eq_corr_ineq_x_1}
\end{equation}
% and
% \begin{equation}
%     c_1+c_2+c_3\ge 1
% \end{equation}
That means that at least two correlations are non-negative among the $c_1,c_2,c_3$.  \\\\
Writing Eqs.
~\eqref{app:eq_var_corr_x_short_a},
~\eqref{app:eq_var_corr_x_short_b} and
~\eqref{app:eq_var_corr_x_short_c}
in a matrix form, with
$\textbf{C}\coloneqq\left[\begin{matrix} 1&-c_3&-c_2\\-c_3&1&-c_1\\-c_2&-c_1&1\end{matrix}\right]$,
and ${\bm \sigma}\coloneqq\left[
\begin{matrix}\sigma_1\\\sigma_2\\ \sigma_3\end{matrix}\right]$,
we have, $\textbf{C}{\bm \sigma}=0$, and hence $\det\textbf{C}=0$, which gives
\begin{equation}
c_1^2+c_2^2+c_3^2+2c_1c_2c_3=1.
\label{app_eq_x_corr_constraint}
\end{equation}
This is exactly the relation between the cosines of the angles in a Euclidean triangle,
\begin{equation}
\cos^2\alpha_1+\cos^2\alpha_2+\cos^2\alpha_3+2\cos\alpha_1\cos\alpha_2\cos\alpha_3=1,
\end{equation}
when $\alpha_1+\alpha_2+\alpha_3=\pi$, fitting the cosine law structure of Eq.~\eqref{app:eq_var_cov_x_1}.
Focusing on the sum and the product of the correlations, by symmetry considerations, and using Eq.~\eqref{app:eq_corr_ineq_2}, we have,
\begin{equation}
    1\le c_1+c_2+c_3\le \frac{3}{2}
    \label{app_eq_x_corr_sum}
\end{equation}
and
\begin{equation}
    -1\le c_1c_2c_3\le \frac{1}{8}.
        \label{app_eq_x_corr_prod}
\end{equation}
% also known to be true for triangle cosines.
% \begin{equation}
% \begin{aligned}
% \sigma^2_a(\hat{x}_b)=
% \sigma^2_c(\hat{x}_a)+\sigma^2_c(\hat{x}_b)-
% 2\sigma_c(\hat{x}_a)\sigma_c(\hat{x}_b)\text{corr}_c(\hat{x}_a,\hat{x}_b).
% \end{aligned}
% \end{equation}

\subsection{Momentum}
\label{app_subsec_momentum}

The expectation value of any function of the momentum of a certain subsystem, say $K$'s, at the initial instant, is perspective-independent
\begin{equation}
    \Braket{f(\hat{p}_K)}_{I}(0) = 
    \int dP_{\bar I} \Psi_{\bar I}^*(P_{\bar I})
    f(p_K)\Psi_{\bar I}(P_{\bar I}) = 
    \int dP_{\bar J}\psi_{\bar J}^*(P_{\bar J}))f(p_K)\psi_{\bar J}(P_{\bar J})= \Braket{f(\hat{p}_K)}_{J}(0),
    \label{eq:<p_I>_any_short}
\end{equation}
where $I,J\in\mathfrak{N}\setminus\{K\}$.
% That means that the momentum variance and covariance expressions of a certain particle are invariant under the switching perspectives
% \begin{equation}
% \sigma^2_I(\hat{p}_K)(0)=\sigma^2_J(\hat{p}_K)(0),
% \label{app:eq_var_p_equiv}
% \end{equation}
% \begin{equation}
%     \begin{aligned}
% \text{cov}_I(\hat{p}_K,\hat{p}_L)(0)=
% \text{cov}_J(\hat{p}_K,\hat{p}_L)(0)
%     \end{aligned}
% \label{app:eq_cov_p_equiv}
% \end{equation}
This is true for any function $f(\hat{p}_K,\hat{p}_L)$,
\begin{equation}\begin{aligned}
\braket{f(\hat{p}_K,\hat{p}_L)}_I(0)=
\braket{f(\hat{p}_K,\hat{p}_L)}_J(0).
\label{app:eq_f(p_I,p_J)}
\end{aligned}\end{equation}
Hence, the momentum covariance matrix from a certain perspective, say $A$'s, excluding one of the particles, say $B$, introduced in Eq.~\eqref{app_eq_cov_mat_tilde_A_div}, is invariant under switching between $A$ and $B$, namely, 
\begin{equation}
\widetilde{\bm\Sigma}^{(p)}_{(A)\setminus \{B\}}(0)=
\left[\begin{matrix}
\sigma^2_A(\hat{p}_C)(0)&
\text{cov}_A(\hat{p}_{C},\hat{p}_D)(0)&\cdots\\
\text{cov}_A(\hat{p}_{D},\hat{p}_C)(0)&
\sigma^2_A(\hat{p}_D)(0)&\cdots\\
\vdots&\vdots&\ddots
\end{matrix}\right]=
\left[\begin{matrix}
\sigma^2_B(\hat{p}_C)(0)&
\text{cov}_B(\hat{p}_{C},\hat{p}_D)(0)&\cdots\\
\text{cov}_B(\hat{p}_{D},\hat{p}_C)(0)&
\sigma^2_B(\hat{p}_D)(0)&\cdots\\
\vdots&\vdots&\ddots
\end{matrix}\right]=
\widetilde{\bm\Sigma}^{(p)}_{(B)\setminus \{A\}}(0).
\label{app_eq_Sigma_p_red_inv}
\end{equation}

Next, for the transformations relating to different perspectives, let us concentrate on the reciprocal descriptions. Due to the momentum constraint in Eq.~\eqref{eq:momentum_constraint}, we have
\begin{equation}
\Braket{f(\hat{p}_I)}_{a\ne I}(0)=
\Braket{f(-\hat{p}_{\bar I})}_{I}(0)=
\int dP_{\bar I}\psi_{\bar I}^*f\left(-\sum_{K\ne I}p_K\right)\psi_{\bar I}=
\Braket{f\left(-\sum_{K\ne I}\hat{p}_K\right)}_I(0).
\label{app:eq_trans_f(p)}
\end{equation}
% \begin{equation}
% \begin{aligned}
% \braket{\hat{p}_A^2}_{B}=
% &\int dP_{\bar A}\psi_{\bar A}^*\left(-\sum_{K\ne A}p_K\right)^2\psi_{\bar A}=\left<\left(-\sum_{K\ne A}\hat{p}_K\right)^2\right>_A=\sum_{K\ne A}\braket{\hat{p}_K^2}+
% \sum_{K\ne L\ne A} \braket{\hat{p}_K\hat{p}_L}_A
% \end{aligned}
% \label{eq_app:<p_I>_any}
% \end{equation}
The perspective-dependent velocity, defined in Eq.~\eqref{eq_QRF_vel}, then transforms as follows
\begin{equation}
\Braket{\hat{v}_{J(I)}}_I=
\Braket{\frac{\hat{p}_J}{m_J}+\frac{\hat{p}_{\bar I}}{m_I}}_I=
\Braket{\frac{\hat{p}_J}{m_J}-\frac{\hat{p}_{I}}{m_I}}_K=
\Braket{
\frac{\hat{p}_J}{m_J}+
\frac{\hat{p}_{\bar K}}{m_K}-
\frac{\hat{p}_{\bar K}}{m_K}
-\frac{\hat{p}_{I}}{m_I}}_K=
\Braket{\hat{v}_{J(K)}}_K-\Braket{\hat{v}_{I(K)}}_K.
\label{app_eq_vel_trans}
\end{equation}
For the momentum variance reciprocal transformation, we have,
\begin{equation}
    \begin{aligned}
\sigma^2_I(\hat{p}_J)(0)=\sigma^2_J(-\hat{p}_{\bar J})(0)=
\sigma^2_J\left(-\sum_{K\ne J}\hat{p}_K\right)(0)=
\sum_{K,L \ne J}\text{cov}_J(\hat{p}_K,\hat{p}_L)(0)=
\sum_{K\ne J}\sigma^2_J(\hat{p}_K)(0)+
\sum_{K\ne L \ne J}\text{cov}_J(\hat{p}_K,\hat{p}_L)(0).
\label{app_eq_sigma(p)_transf}
    \end{aligned}
\end{equation}
The transformation of the momentum covariance between particles $I$ and $J$ is as follows,
\begin{equation}
\begin{aligned}
\text{cov}_{a\ne I,J}(\hat{p}_I,\hat{p}_J)(0)=
&-\text{cov}_{I}(\hat{p}_{\bar I},\hat{p}_J)(0)=
-\sigma_I^2(\hat{p}_J)(0)
-\sum_{L\ne I,J}\text{cov}_I(\hat{p}_L,\hat{p}_J)(0)=
-\sigma_I^2(\hat{p}_J)(0)-
\text{cov}_I(\hat{p}_{\overline{IJ}},\hat{p}_J)(0)=\\
&-\text{cov}_{J}(\hat{p}_{\bar J},\hat{p}_I)(0)=
-\sigma_J^2(\hat{p}_I)(0)
-\sum_{L\ne I,J}\text{cov}_J(\hat{p}_I,\hat{p}_L)(0)=
-\sigma_J^2(\hat{p}_I)(0)-
\text{cov}_J(\hat{p}_{\overline{IJ}},\hat{p}_I)(0),
\end{aligned}
\label{app:eq_cov_p_trnas}
\end{equation}
where $\hat{p}_{\overline{IJ}}
=\sum_{K\ne I,J}\hat{p}_K$,
is related to the variance of $J$ and the sum of its covariances with the rest of the system from the perspective of $I$, and vice versa. Writing these in terms of variances and correlations, we have,
\begin{equation}
\begin{aligned}
\sigma_I(\hat{p}_J)(0)+
\sigma_J(\hat{p}_I)(0)\text{corr}_{u\ne I,J}(\hat{p}_I,\hat{p}_J)(0)+
\sigma_{I,J}(\hat{p}_{\overline{IJ}})(0)\text{corr}_I(\hat{p}_{\overline{IJ}},\hat{p}_J)(0)=0,
\end{aligned}
\label{app:eq_var_corr_long_1}
\end{equation}
and,
\begin{equation}
\begin{aligned}
\sigma_J(\hat{p}_I)(0)+
\sigma_I(\hat{p}_J)(0)
\text{corr}_{u\ne I,J}(\hat{p}_I,\hat{p}_J)(0)+
\sigma_{I,J}(\hat{p}_{\overline{IJ}})(0)
\text{corr}_J(\hat{p}_{\overline{IJ}},\hat{p}_I)(0)=0.
\end{aligned}
\label{app:eq_var_corr_long_2}
\end{equation}
% In a general case of $\mathscr{N}$ particles, 
Adopting for simplicity, the shorthand notations:\\
\begin{equation}
    \begin{aligned}
&c_1\equiv
\text{corr}_I(\hat{p}_{\overline{IJ}},\hat{p}_J)(0),\hspace{5mm} 
c_2\equiv
\text{corr}_J(\hat{p}_{\overline{IJ}},\hat{p}_I)(0),\hspace{5mm}
c_3\equiv
\text{corr}_{u\ne I,J}(\hat{p}_I,\hat{p}_J)(0),
    \end{aligned}
\end{equation}
and,
\begin{equation}
\sigma_1\equiv\sigma_I(\hat{p}_J)(0),\hspace{5mm}
\sigma_2\equiv\sigma_J(\hat{p}_I)(0), \hspace{5mm}
\sigma_3\equiv\sigma_{I,J}(\hat{p}_{\overline{IJ}})(0),
\end{equation}
we may rewrite the Eqs.~\eqref{app:eq_var_corr_long_1} and~\eqref{app:eq_var_corr_long_2} as follows,
\begin{equation}\begin{aligned}
\sigma_1+\sigma_2c_3+\sigma_3 c_1=0,
\label{app:eq_var_corr_short_1}
\end{aligned}\end{equation} 
\begin{equation}\begin{aligned}
\sigma_2+\sigma_1c_3+\sigma_3 c_2=0.
\label{app:eq_var_corr_short_2}
\end{aligned}\end{equation} 
This gives the following variances inequalities,
\begin{equation}\begin{aligned}
\sigma_1\le\sigma_2+\sigma_3,
\hspace{5mm}
\sigma_2\le\sigma_1+\sigma_3,
\end{aligned}\end{equation} 
which is equivalent to, 
\begin{equation}\begin{aligned}
|\sigma_1-\sigma_2|\le\sigma_3.
\end{aligned}\end{equation} 
Next, by adding the Eqs.~\eqref{app:eq_var_corr_short_1} and~\eqref{app:eq_var_corr_short_2}, we have 
\begin{equation}\begin{aligned}
(\sigma_1+\sigma_2)(1+c_3)+\sigma_3(c_1+c_2)=0,
\end{aligned}\end{equation}
which gives the following correlations inequality
\begin{equation}\begin{aligned}
c_1+c_2\le0
\label{app:eq_c_1+c_2}.
\end{aligned}\end{equation} 
% and
% \begin{equation}\begin{aligned}
% (\sigma_1-\sigma_2)(1-c_3)+\sigma_3(c_1-c_2)=0
% \end{aligned}\end{equation}
% from which follows that
% \begin{equation}\begin{aligned}
% |c_1-c_2|\le 1-c_3
% \end{aligned}\end{equation} 
Writing Eqs.~\eqref{app:eq_var_corr_short_1} and~\eqref{app:eq_var_corr_short_2} in the matrix form, one has,
\begin{equation}\begin{aligned}
\left[
\begin{matrix}
1&c_3\\c_3&1
\end{matrix}
\right]
\left[
\begin{matrix}
\sigma_1\\ \sigma_2
\end{matrix}
\right]=-\sigma_3
\left[
\begin{matrix}
c_1\\ c_2
\end{matrix}
\right].
\end{aligned}\end{equation} 
Multiplying it from both sides with 
$
\left[
\begin{matrix}
1&c_3\\c_3&1
\end{matrix}
\right]^{-1}=
\frac{1}{1-c_3^2}
\left[
\begin{matrix}
1&-c_3\\-c_3&1
\end{matrix}
\right]
$, we have,
\begin{equation}\begin{aligned}
\left[
\begin{matrix}
\sigma_1\\ \sigma_2
\end{matrix}
\right]=
\sigma_3
\frac{1}{1-c_3^2}
\left[
\begin{matrix}
-1&c_3\\c_3&-1
\end{matrix}
\right]
\left[
\begin{matrix}
c_1\\ c_2
\end{matrix}
\right].
\end{aligned}\end{equation} 
For $\sigma_1$ and $\sigma_2$ we have,
\begin{equation}\begin{aligned}
\sigma_i=\frac{\bar\sigma(c_3c_j-c_i)}{1-c_3^2},
\end{aligned}\end{equation} 
where, $i,j=1,2$, which gives
\begin{equation}\begin{aligned}
c_i\le c_3 c_j.
\end{aligned}\end{equation}

\section{Invariance of the determinant of the total covariance matrix under changing perspectives}
\label{app:CM_inv_under_QRF_chng}
% Since the determinant of the perspective-dependent position-momentum covariance matrix for a quadratic Hamiltonians is constant in time. as we showed in Subsection~\ref{app_subsec:det_cov_time_ind}, we may concentrate for simplicity on the initial expressions.
In this section, we present the transformations between different QRFs of the total, perspective-dependent, position-momentum covariance matrix of all the particles in a certain QRF. More precisely, we will show that its determinant is invariant when changing perspectives.
In the following, we will focus on the vector of $2(\mathscr{N}-1)$ position and momentum operators from a certain perspective, say $A$'s, of the form, 
$\hat{\textbf{R}}^T_{\bar A}=
\left[ \begin{matrix} 
\hat{\textbf{X}}^T_{\bar A}&\hat{\textbf{P}}^T_{\bar A}
\end{matrix}\right]$, where 
$\hat{\textbf{X}}^T_{\bar A}=
\left[ \begin{matrix} 
\hat{x}_B&\hat{x}_C&\ldots&\hat{x}_\mathscr{N}
\end{matrix}\right]$, and 
$\hat{\textbf{P}}^T_{\bar A}=
\left[ \begin{matrix} 
\hat{p}_B&\hat{p}_C&\ldots&\hat{p}_\mathscr{N}
\end{matrix}\right]$. 
Using Eqs.~\eqref{app:eq_trans_f(x)} and~\eqref{app:eq_trans_f(p)}, we may write the following transformations,
\begin{equation}
    \begin{aligned}
&\Braket{\hat{\textbf{X}}_{\bar A}^T}_A(0)=
\Braket{ \left[ \begin{matrix} 
\hat{x}_B&\hat{x}_C&\hat{x}_D&\cdots&
\end{matrix}\right]}_A(0)=
\Braket{\left[\begin{matrix}
-\hat{x}_A&\hat{x}_C-\hat{x}_A&\hat{x}_C-\hat{x}_A&\cdots&
\end{matrix}\right]}_B(0)=\\
&\Braket{\left[\begin{matrix}
\hat{x}_A&\hat{x}_C&\hat{x}_D&\cdots&
\end{matrix}\right]}_B(0)
\left[\begin{matrix}
    -1&-1&-1&\cdots\\
    0&1&0&\cdots\\
    0&0&1&\cdots\\
    \vdots&\vdots&\vdots&\ddots
\end{matrix}\right]=
\Braket{\hat{\textbf{X}}_{\bar B}^T}_B(0)
{\bm \alpha}_{BA}^T,
    \end{aligned}
\end{equation}
and 
\begin{equation}
    \begin{aligned}
&\Braket{\hat{\textbf{P}}_{\bar A}^T}_A(0)=
\Braket{ \left[ \begin{matrix} 
\hat{p}_B&\hat{p}_C&\hat{p}_D&\cdots&
\end{matrix}\right]}_A(0)=
\Braket{\left[\begin{matrix}
-\sum_{I\ne B}\hat{p}_I&\hat{p}_C&\hat{p}_C&\cdots&
\end{matrix}\right]}_B(0)=\\
&\Braket{\left[\begin{matrix}
\hat{p}_A&\hat{p}_C&\hat{p}_D&\cdots&
\end{matrix}\right]}_B(0)
\left[\begin{matrix}
    -1&0&0&\cdots\\
    -1&1&0&\cdots\\
    -1&0&1&\cdots\\
    \vdots&\vdots&\vdots&\ddots
\end{matrix}\right]=
\Braket{\hat{\textbf{P}}_{\bar B}^T}_B(0)
{\bm \beta}_{BA}^T,
    \end{aligned}
\end{equation}
where
$ 
{\bm \beta}_{BA}={\bm \alpha}_{BA}^T=
\left[\begin{matrix}
    -1&-1&-1&\cdots\\
    0&1&0&\cdots\\
    0&0&1&\cdots\\
    \vdots&\vdots&\vdots&\ddots
\end{matrix}\right]
$. Summarizing the above, we have, 
\begin{equation}
\Braket{\hat{\textbf{X}}_{\bar A}}_A(0)=
{\bm \alpha}_{AB}
\Braket{\hat{\textbf{X}}_{\bar B}}_B(0),
\hspace{5mm}
\Braket{\hat{\textbf{X}}_{\bar A}^T}_A(0)=
\Braket{\hat{\textbf{X}}_{\bar B}^T}_B(0) 
{\bm \alpha}_{AB}^T,
\hspace{5mm}
\Braket{\hat{\textbf{X}}_{\bar A}\hat{\textbf{X}}_{\bar A}^T}_A(0)=
{\bm \alpha}_{AB}
\Braket{\hat{\textbf{X}}_{\bar B}\hat{\textbf{X}}_{\bar B}^T}_B(0){\bm \alpha}_{AB}^T,
\end{equation}
\begin{equation}
\Braket{\hat{\textbf{P}}_{\bar A}}_A(0)=
{\bm \beta}_{AB}
\Braket{\hat{\textbf{P}}_{\bar B}}_B(0),
\hspace{5mm}
\Braket{\hat{\textbf{P}}_{\bar A}^T}_A(0)=
\Braket{\hat{\textbf{P}}_{\bar B}^T}_B(0)
{\bm \beta}_{AB}^T,
\hspace{5mm}
\Braket{\hat{\textbf{P}}_{\bar A}\hat{\textbf{P}}_{\bar A}^T}_A(0)=
{\bm \beta}_{AB}
\Braket{\hat{\textbf{P}}_{\bar B}\hat{\textbf{P}}_{\bar B}^T}_B(0)
{\bm \beta}_{AB}^T,
\end{equation}
and,
\begin{equation}
\Braket{\hat{\textbf{R}}_{\bar A}}_A(0)=
{\bm \Theta}_{AB}
\Braket{\hat{\textbf{R}}_{\bar B}}_B(0),
\hspace{5mm}
\Braket{\hat{\textbf{R}}_{\bar A}^T}_A(0)=
\Braket{\hat{\textbf{R}}_{\bar B}^T}_B(0)
{\bm \Theta}_{AB}^T,
\hspace{5mm}
\Braket{\hat{\textbf{R}}_{\bar A}
\hat{\textbf{R}}_{\bar A}^T}_A(0)=
{\bm \Theta}_{AB}
\Braket{\hat{\textbf{R}}_{\bar B}\hat{\textbf{R}}_{\bar B}^T}_B(0)
{\bm \Theta}_{AB}^T,
\label{app_eq_RR^T_trans}
\end{equation}
where, 
${\bm \alpha}_{AB}\equiv{\bm \alpha}_{BA}$, ${\bm \beta}_{AB}\equiv {\bm \beta}_{BA}$, and,
${\bm \Theta}_{BA}=
{\bm \Theta}_{AB}=\left[\begin{matrix}
{\bm \alpha}_{AB}&0\\0&{\bm \beta}_{AB}
\end{matrix}\right]$. \\\\
Using Eq.~\eqref{app_eq_RR^T_trans} the covariance matrix, introduced in Eq.~\eqref{eq_Sigma_tilde}, transforms as follows,
\begin{equation}
    \begin{aligned}
\widetilde{\bm{\Sigma}}^{(x,p)}_{(A)}(0)=
&\bm{\Theta}_{AB}
\widetilde{\bm{\Sigma}}^{(x,p)}_{(B)}(0)
{\bm \Theta}_{AB}^T.
    \end{aligned}
\end{equation}
Since,
\begin{equation}
    \det{\bm \Theta}_{AB}=\det{\bm \Theta}_{AB}^T=
\det{\bm \theta}_{AB}\det{\bm \theta}_{AB}^T=1,
\end{equation}
the determinant of the position-momentum covariance matrix is perspective invariant,
\begin{equation}
\det\widetilde{\bm{\Sigma}}^{(x,p)}_{(A)}(0)=
\det\widetilde{\bm{\Sigma}}^{(x,p)}_{(B)}(0).
\end{equation}
This is true for the position and momentum covariances separately,
\begin{equation}
\det \widetilde{\bm{\Sigma}}^{(r)}_{(A)}(0)=
\det \widetilde{\bm{\Sigma}}^{(r)}_{(B)}(0),
\label{app_eq_Sigma_r_inv}
\end{equation}
where $r=x,p$, since 
$\det \bm{\alpha}_{AB}\det \bm{\alpha}_{AB}^T=
\det \bm{\beta}_{AB}\det \bm{\beta}_{AB}^T$=1.\\

Since the determinant is unchanged under switching rows and columns, this result still holds for a different, and, sometimes more convenient arrangement of operators in a vector, 
$\hat{\textbf{R}}_{\bar A}=
\left[ \begin{matrix} 
\hat{x}_B&\hat{p}_B&\hat{x}_C &\hat{p}_C&\ldots&
\end{matrix}\right]$, giving,
\begin{equation}
\det{\bm{\Sigma}}^{(x,p)}_{(A)}(0)=
\det{\bm{\Sigma}}^{(x,p)}_{(B)}(0).
\label{app_eq_det_Sigma_inv}
\end{equation}

\section{Time evolution of the position and momentum second moments without interactions}
\label{app:section_QRF_UR_no_interactions}
We start by observing that the expectation value of the momentum of a certain particle, say $J$, in a certain frame, say $I$, is time-independent, $e^{it H_{\bar{I}}}\psi_{\bar I}^*(P_{\bar I}) p_J e^{-it H_{\bar{I}}} \psi_{\bar I}(P_{\bar I}) = \psi_{\bar I}^*(P_{\bar I})p_J\psi_{\bar I}(P_{\bar I})$, giving
\begin{equation}
\braket{f(\hat{p}_J)}_{I}(t)=\int dP_{\bar I} \psi_{\bar I}^*(P_{\bar I})
f(p_J)\psi_{\bar I}(P_{\bar I})=\braket{f(\hat{p}_J)}_{I}(0),
\label{eq:<p_I>_time_ind}
\end{equation}
where $I\in\mathfrak{N}\setminus\{J\}$.
That means that the momentum variance and covariance expressions are time-independent,
\begin{equation}
\sigma^2_I(\hat{p}_J)(t)=\sigma^2_I(\hat{p}_J)(0),
\label{app:eq_sigma_p_inv}
\end{equation}
and
\begin{equation}
    \begin{aligned}
\text{cov}_I(\hat{p}_J,\hat{p}_K)(t)=
\text{cov}_I(\hat{p}_J,\hat{p}_K)(0).
    \end{aligned}
\label{app_eq:cov_p_const}
\end{equation}
Hence, using the above and Eq.~\eqref{app_eq_Sigma_p_red_inv}, we see that the total momentum covariance matrices from the perspective of any two QRFs, excluding each other, coincide for any $t$ and $t'$,
\begin{equation}
\widetilde{\bm{\Sigma}}_{(I)\setminus \{J\}}^{(p)}(t)= \widetilde{\bm{\Sigma}}_{(J)\setminus \{I\}}^{(p)}(t').
\label{app_eq_momentum_CM_time_indep}
\end{equation}
For a general function of the position operator $\hat{x}_J$ in the perspective of particles $I$ we have the following expectation value,
\begin{equation}\begin{aligned}
&\braket{f(\hat x_J)}_I(t)=
\langle\psi_{\bar I}(t)|\hat x_J|\psi_{\bar I}(t)\rangle=
\int dP_{\bar I}
\psi_{\bar I}^*(P_{\bar I})e^{itH_{\bar I}}
f\left(i\frac{d}{dp_J}\right) e^{-itH_{\bar I}} \psi_{\bar I}(P_{\bar I}) =\\
&=\int dP_{\bar I}
\psi_{\bar I}^*(P_{\bar I})
f\left(i\frac{d}{dp_J}+ tv_{J(I)}\right)
\psi_{\bar I}(P_{\bar I})
=\Braket{f\left(\hat x_J+t\hat{v}_{J(I)}\right)}_I(0),
\label{eq:<x_I>_J(t_J)}
\end{aligned}\end{equation} 
where we used the derivative of the perspective-dependent Hamiltonian,
\begin{equation}\begin{aligned}
\frac{d}{dp_J} H_{\bar I}=\frac{d}{dp_J}
\left( \frac{ p_{\bar{I}}^2}{2m_I} + \sum_{K\ne I} \frac{{p}_K^2}{2m_K}
\right)=\frac{p_J}{m_J}+\frac{ p_{\bar{I}}}{m_I}
\equiv v_{J(I)},
\label{app:dH_J/dp_I}
\end{aligned}\end{equation} 
which reflects the relative velocity operator,
\begin{equation}
    \hat{v}_{J(I)}=
    \frac{\hat{p}_J}{m_J}+\frac{\hat{p}_{\bar I}}{m_I}.
    \label{eq:rel_velocity}
\end{equation}
% \begin{equation}
% \braket{\hat{v}_{IJ}}_K=\braket{\hat{v}_{IJ}}_L=\braket{(\hat{v}_{IJ})_{\bar J}}_J
% \label{app:eq_(v_I|J)=(v_IJ)_K}
% \end{equation}
% \begin{equation}
%     \braket{\hat{v}_{I\vert J}}_J=-\braket{\hat{v}_{J\vert I}}_I
% \end{equation}
The position-momentum covariance evolution may be obtained,
\begin{equation}\begin{aligned}
\text{cov}_I\left(\hat{x}_J,\hat{p}_K\right)(t)=
&\text{cov}_I\left(\hat{x}_J+t\hat{v}_{J(I)},\hat{p}_K\right)(0)=
\text{cov}_I\left(\hat{x}_J,\hat{p}_K\right)(0)+
t\text{cov}_I\left(\hat{v}_{J(I)},\hat{p}_K\right)(0).
\label{app:eq_cov_I(x_J,p_K)(t)}
\end{aligned}\end{equation}
The position-position covariance evolution is of the form,
\begin{equation}\begin{aligned}
\text{cov}_I(\hat{x}_J,\hat{x}_K)(t)=
&\text{cov}_I\left(\hat{x}_J+t\hat{v}_{J(I)},\hat{x}_K+t\hat{v}_{K(I)}\right)(0)=\\
&\text{cov}_I\left(\hat{x}_J,\hat{x}_K\right)(0)+
t\Big[
\text{cov}_I\left(\hat{x}_J,\hat{v}_{K(I)}\right)(0)+
\text{cov}_I\left(\hat{v}_{J(I)},\hat{x}_K\right)(0)
\Big]+
t^2\text{cov}_I\left(\hat{v}_{J(I)},\hat{v}_{K(I)}\right)(0),
\label{app:eq_cov_I(x_J,x_K)(t)}
\end{aligned}\end{equation} 
where, using Eq.~\eqref{eq:rel_velocity} we have
\begin{equation}\begin{aligned}
\text{cov}_I\left(\hat{v}_{J(I)},\hat{p}_K\right)(0)=
&\text{cov}_I\left(
\frac{\hat{p}_J}{m_J}+\frac{\sum_{L\ne I}\hat{p}_L}{m_I},\hat{p}_K
\right)(0)=
\frac{1}{m_J}\text{cov}_I\left(\hat{p}_J,\hat{p}_K\right)(0)+
\frac{1}{m_I}\sum_{L\ne I} \text{cov}_I\left(\hat{p}_L,\hat{p}_K\right)(0),
\label{app:eq_cov_I(v_J,p_K)(0)}
\end{aligned}\end{equation} 
\begin{equation}
    \begin{aligned}
\text{cov}_I\left(\hat{x}_J,\hat{v}_{K(I)}\right)(0)=
\text{cov}_I\left(\hat{x}_J,\frac{\hat{p}_K}{m_K}+\sum_{L\ne I}\frac{\hat{p}_L}{m_I}\right)(0)=
\frac{1}{m_K}\text{cov}_I(\hat{x}_J,\hat{p}_K)_I(0)+
\frac{1}{m_I}\sum_{L\ne I}
\text{cov}_I(\hat{x}_J,\hat{p}_L)_I(0),
\end{aligned}
\label{app:eq_cov_I(x_J,v_K)(0)}
\end{equation} 
and,
\begin{equation}\begin{aligned}
&\text{cov}_I\left(\hat{v}_{J(I)},\hat{v}_{K(I)}\right)=
\text{cov}_I\left(
\frac{\hat{p}_J}{m_J}+\frac{\sum_{L\ne I}\hat{p}_L}{m_I},
\frac{\hat{p}_K}{m_K}+\frac{\sum_{L\ne I}\hat{p}_L}{m_I}
\right)=\\
&\frac{1}{m_Jm_K}\text{cov}_I(\hat{p}_J,\hat{p}_K)+
\frac{1}{m_Jm_I}\sum_{L\ne I}\text{cov}_I(\hat{p}_J,\hat{p}_L)+
\frac{1}{m_Km_I}\sum_{L\ne I}\text{cov}_I(\hat{p}_K,\hat{p}_L)+
\frac{1}{m_I^2}\sum_{L,M\ne I}\text{cov}_I(\hat{p}_L,\hat{p}_M).
\label{app:eq_cov_I(v_J,v_K)(0)}
\end{aligned}\end{equation}

\section{Time independence of the covariance matrix determinant of the total system in the case of quadratic Hamiltonian}\label{app_subsec:det_cov_time_ind}
In the following, we consider the perspective-dependent Hamiltonian in Eq.~\eqref{eq:Hamiltonian_persp_dep} in a special case of quadratic Hamiltonian of the form,
\begin{equation}\begin{aligned}
\hat{H}_{\bar A}=\frac{1}{2}\hat{\textbf{R}}_{\bar A}^T\textbf{G}_{\bar A}\hat{\textbf{R}}_{\bar A},
\end{aligned}\end{equation} 
where, $\hat{\textbf{R}}_{\bar A}^T=
\left[\begin{matrix}
\hat{x}_B&\hat{p}_B&\hat{x}_C&\hat{p}_C&\cdots&\hat{x}_\mathscr{N}&\hat{p}_\mathscr{N}
\end{matrix}\right]$, and, $\textbf{G}_{\bar A}$ is a symmetric $2(\mathscr{N}-1)\times2(\mathscr{N}-1)$ matrix. 
Frame-dependent position-momentum covariance matrix introduced in Eq.~\eqref{eq:cov_mat_A} is of the form,
\begin{equation}
    \bm{\Sigma}^{(x,p)}_{(A)}(t)=
\frac{1}{2}
\left(
\Braket{\hat{\textbf{R}}_{\bar A}\hat{\textbf{R}}_{\bar A}^T}_A(t)+
\Braket{\hat{\textbf{R}}_{\bar A}\hat{\textbf{R}}_{\bar A}^T}_A^*(t)
\right)-
\Braket{\hat{\textbf{R}}_{\bar A}}_A(t)
\Braket{\hat{\textbf{R}}_{\bar A}^T}_A(t).
\label{eq:cov=<RR^T>}
\end{equation}
% where,
% \begin{equation}\begin{aligned}
% \Braket{\hat{\textbf{R}}_{\bar A}}_A(t)=
% \Braket{\psi_{\bar A}(t)|\hat{\textbf{R}}_{\bar A}|\psi_{\bar A}(t)}=
% \Braket{\psi_{\bar A}(0)|U_{\bar A}^T(t)\hat{\textbf{R}}_{\bar A}U_{\bar A}(t)|\psi_{\bar A}(0)}
% \end{aligned}\end{equation} 
The Heisenberg equations for each term are as follows,
\begin{equation}\begin{aligned}
\frac{d}{dt}\Braket{\hat{\textbf{R}}_{\bar A}}_A=
i\Braket{\left[\hat{H}_{\bar A},\hat{\textbf{R}}_{\bar A}\right]}_A,
\label{eq:d/dt<R>}
\end{aligned}\end{equation} 
and,
\begin{equation}\begin{aligned}
\frac{d}{dt}\Braket{\hat{\textbf{R}}_{\bar A}\hat{\textbf{R}}_{\bar A}^T}_A=
i\Braket{\left[\hat{H}_{\bar A},\hat{\textbf{R}}_{\bar A}\hat{\textbf{R}}_{\bar A}^T\right]}_A.
\label{eq:d/dt<RR^T>}
\end{aligned}\end{equation} 
Let us start the calculation of the commutation relations in the above with individual components,
\begin{equation}
\begin{aligned}
\left(\hat{R}_{\bar A\,i}G_{\bar A\,ij}\hat{R}_{\bar A\,j}\right) \hat{R}_{\bar A\,k}=
&
\hat{R}_{\bar A\,i}G_{\bar A\,ij}\hat{R}_{\bar A\,k} \hat{R}_{\bar A\,j}-
\hat{R}_{\bar A\,i}G_{\bar A\,ij}
\left[\hat{R}_{\bar A\,k} \hat{R}_{\bar A\,j}\right]=\\
&\hat{R}_{\bar A\,k}\hat{R}_{\bar A\,i}G_{\bar A\,ij} \hat{R}_{\bar A\,j}-
\left[\hat{R}_{\bar A\,k},\hat{R}_{\bar A\,i}\right]
G_{\bar A\,ij} \hat{R}_{\bar A\,j}-
\hat{R}_{\bar A\,i}G_{\bar A\,ij}
\left[\hat{R}_{\bar A\,k} \hat{R}_{\bar A\,j}\right]=\\
&\hat{R}_{\bar A\,k}\hat{R}_{\bar A\,i}G_{\bar A\,ij} \hat{R}_{\bar A\,j}-2i\Omega_{ki}G_{\bar A\,ij} \hat{R}_{\bar A\,j},
\end{aligned}
\end{equation}
where we denoted, $\Omega_{kj}=-i\left[\hat{R}_{\bar A\,k} \hat{R}_{\bar A\,j}\right]=\left(
\bigoplus_{i=1}^\mathscr{N}
\left[\begin{matrix}
    0 & 1\\-1&0
\end{matrix}\right]\right)_{kj}$, and used the fact that $\textbf{G}_{\bar A}$ is symmetric. That means that we have,
\begin{equation}
\left[\hat{H}_{\bar A},\hat{\textbf{R}}_{\bar A}\right]=-i\bm{\Omega}\textbf{G}_{\bar A}\hat{\textbf{R}}_{\bar A},
\end{equation}
which gives the $\Braket{\hat{\textbf{R}}_{\bar A}}_A$ and $\Braket{\hat{\textbf{R}}_{\bar A}^T}_A$ time dependence in Eq.~\eqref{eq:d/dt<R>},
\begin{equation}
\frac{d}{dt}\Braket{\hat{\textbf{R}}_{\bar A}}_A=
i\Braket{\left[\hat{H}_{\bar A},\hat{\textbf{R}}_{\bar A}\right]}_A=
{\bm\Omega}\textbf{G}_{\bar A}\Braket{\hat{\textbf{R}}_{\bar A}}_A,
\end{equation} 
and, 
\begin{equation}
\frac{d}{dt}\Braket{\hat{\textbf{R}}_{\bar A}^T}_A=
i\Braket{\left[\hat{H}_{\bar A},\hat{\textbf{R}}_{\bar A}^T\right]}_A=
\Braket{\hat{\textbf{R}}_{\bar A}^T}_A
\textbf{G}_{\bar A}^T{\bm\Omega}^T.
\end{equation} 
Hence, we have,
\begin{equation}\begin{aligned}
\Braket{\hat{\textbf{R}}_{\bar A}}_A(t)=S_{\bar A}(t)
\Braket{\hat{\textbf{R}}_{\bar A}}_A(0),
\end{aligned}\end{equation} 
and,
\begin{equation}\begin{aligned}
\Braket{\hat{\textbf{R}}_{\bar A}^T}_A(t)=
\Braket{\hat{\textbf{R}}_{\bar A}^T}_A(0)S_{\bar A}^T(t),
\end{aligned}\end{equation} 
where 
$\textbf{S}_{\bar A}(t)=e^{{\bm\Omega\textbf{G}}_{\bar A}t}$, 
$\textbf{S}_{\bar A}^T(t)=e^{-{\textbf{G}}_{\bar A}\bm\Omega t}$. 
The matrix $\textbf{S}(t)$ is a symplectic matrix since it satisfies 
$\textbf{S}{\bm \Omega}\textbf{S}^T=
\textbf{S}^T{\bm \Omega}\textbf{S}={\bm \Omega}$. The Heisenberg equation in Eq.~\eqref{eq:d/dt<RR^T>} may be obtained in a similar way, 
\begin{equation}\begin{aligned}
\hat{H}_{\bar A}\hat{\textbf{R}}_{\bar A}\hat{\textbf{R}}_{\bar A}^T=
\hat{\textbf{R}}_{\bar A}\hat{H}_{\bar A}\hat{\textbf{R}}_{\bar A}^T
-i{\bm\Omega}\textbf{G}_{\bar A}\hat{\textbf{R}}_{\bar A}
\hat{\textbf{R}}_{\bar A}^T=
\hat{H}_{\bar A}\hat{\textbf{R}}_{\bar A}\hat{\textbf{R}}_{\bar A}^T
-i\hat{\textbf{R}}_{\bar A}\hat{\textbf{R}}_{\bar A}^T \textbf{G}_{\bar A}^T{\bm\Omega}^T
-i{\bm\Omega}\textbf{G}_{\bar A}\hat{\textbf{R}}_{\bar A}\hat{\textbf{R}}_{\bar A}^T,
\end{aligned}\end{equation} 
giving,
\begin{equation}\begin{aligned}
\frac{d}{dt}\Braket{\hat{\textbf{R}}_{\bar A}\hat{\textbf{R}}_{\bar A}^T}_A=
i
\Braket{
\left[\hat{H}_{\bar A},\hat{\textbf{R}}_{\bar A}\hat{\textbf{R}}_{\bar A}^T\right]}_A=
\Braket{\hat{\textbf{R}}_{\bar A}\hat{\textbf{R}}_{\bar A}^T}_A
\textbf{G}_{\bar A}^T {\bm\Omega}^T
+{\bm\Omega}\textbf{G}_{\bar A}
\Braket{\hat{\textbf{R}}_{\bar A}\hat{\textbf{R}}_{\bar A}^T}_A,
\end{aligned}\end{equation} 
and,
\begin{equation}\begin{aligned}
\Braket{\hat{\textbf{R}}_{\bar A}\hat{\textbf{R}}_{\bar A}^T}_A(t)=
S_{\bar A}(t)
\Braket{\hat{\textbf{R}}_{\bar A}\hat{\textbf{R}}_{\bar A}^T}_A(0)
S_{\bar A}^T(t)
\end{aligned}\end{equation} 
Hence, the time dependence of the covariance matrix in Eq.~\eqref{eq:cov=<RR^T>} is of the form:
\begin{equation}\begin{aligned}
\bm{\Sigma}_{(A)}^{(x,p)}(t)= S_{\bar A}(t) \bm{\Sigma}_{(A)}^{(x,p)}(0) S_{\bar A}^T(t)
\end{aligned}\end{equation} 
And, since
\begin{equation}\begin{aligned}
\det S_{\bar A}(t)\det S_{\bar A}^T(t)=1
\end{aligned}\end{equation} 
the determinant of the covariance matrix is constant in this case
\begin{equation}\begin{aligned}
\det\bm{\Sigma}_{(A)}^{(x,p)}(t)=
\det\bm{\Sigma}_{(A)}^{(x,p)}(0)
\label{app:cov_QRF_time_indep}
\end{aligned}\end{equation} 
% The case without interactions is a special case of the quadratic Hamiltonian with the $\textbf{G}$ matrix of the form
% \begin{equation}
%     \textbf{G}=
% \end{equation}

\bibliography{qrf_bib}
\end{document}